\documentclass{aa}  

\usepackage{graphicx}
\usepackage{lscape}
\usepackage{natbib}
\bibpunct{(}{)}{;}{a}{}{,} 
\newcommand\msun{\rm\,M_\odot}

\usepackage{txfonts}
\begin{document}

   \title{The colour-magnitude relation of globular clusters in Centaurus and Hydra}

   \subtitle{Constraints on star cluster self-enrichment with a link to massive Milky Way GCs}

   \author{J. Fensch\inst{1} \inst{2}
          \and
          S. Mieske\inst{2}
          \and
          J. M\"uller-Seidlitz\inst{3}
          \and
          M. Hilker\inst{3}
          }

   \institute{Ecole polytechnique, Route de Saclay, 91128, Palaiseau, France
                      \and
            European Southern Observatory, Alonso de Cordova 3107, Vitacura, Santiago de Chile, Chile
            \and
            European Southern Observatory, Karl-Schwarzschild-Str. 2, 85748 Garching b. M\"unchen, Germany
             }

   \date{}

 
  \abstract
   {} 
   {We investigate the colour-magnitude relation of metal-poor globular clusters, the so called blue tilt, in the Hydra and Centaurus galaxy clusters and constrain the primordial conditions for star cluster self-enrichment.}
{We analyse U,I photometry for about 2500 globular clusters in the
  central regions of Hydra and Centaurus, based on FORS1@VLT data. We
  measure the relation between mean colour and luminosity for the blue
  and red subpopulation of the globular cluster samples. We convert
  these relations into mass-metallicity space and compare the
  obtained GC mass-metallicity relation with predictions from the star
  cluster self-enrichment model by Bailin \& Harris (2009). For this
  we include effects of dynamical and stellar evolution and a
  physically well motivated primordial mass-radius scaling.}
{We obtain a mass-metallicity scaling of $Z \propto M^{0.27 \pm 0.05}$
  for Centaurus GCs and $Z \propto M^{0.40 \pm 0.06}$ for Hydra GCs,
  consistent with the range of observed relations in other
  environments. We find that the GC mass-metallicity
  relation already sets in at present-day masses of a few $10^5 \msun$
  and is well established in the luminosity range of massive MW
  clusters like omega Centauri. The inclusion of a primordial
  mass-radius scaling of star clusters significantly improves the fit
  of the self-enrichment model to the data. The self-enrichment model accurately reproduces the observed relations for average primordial
  half-light radii $r_h \sim 1-1.5$ pc, star formation efficiencies
  $f_\star \sim 0.3-0.4$, and pre-enrichment levels of $[Fe/H] ~ -1.7$
  dex. The slightly steeper blue tilt for Hydra can be explained
  either by a $\sim$30\% smaller average $r_h$ at fixed $f_\star \sim
  0.3$, or analogously by a $\sim$20\% smaller $f_\star$ at fixed $r_h
  \sim 1.5$ pc Within the self-enrichment scenario, the observed blue
  tilt implies a correlation between GC mass and width of the stellar
  metallicity distribution. We find that this implied correlation
  matches the trend of width with GC mass measured in Galactic GCs,
  including extreme cases like Omega Cen and M54. }
{ First, we found that a  primordial star cluster mass-radius relation provides a
  significant improvement to the self-enrichment model
  fits. Second we show that broadened metallicity distributions as found in some
  massive MW globular clusters may have arisen naturally from
  self-enrichment processes, without the need of a dwarf galaxy
  progenitor.}


   \maketitle
%

\section{Introduction}

\subsection{Colour magnitude relation of extragalactic globular clusters}

The broadband colour distribution of globular clusters (GCs) exhibits
a strong bimodality especially in globular cluster systems of massive
galaxies (e.g. Gebhardt \& Kissler-Patig~\citeyear{Gebhar99}; Kundu \&
Whitmore~\citeyear{Kundu01}, Larsen et al.~\citeyear{Larsen01}), which
is often interpreted as a result of star formation and galaxy assembly
processes in the early universe (see the review of Brodie \&
Strader~\citeyear{Brodie06}). In the last decade, Hubble Space Telescope (HST) and ground-based
wide-field imaging of hundreds of GC systems has provided uswith very detailed information on bimodality (e.g. Dirsch et
al.~\citeyear{Dirsch03}, Bassino et al.~\citeyear{Bassin06}, Peng et
al.~\citeyear{Peng06}, Harris et al.~\citeyear{Harris06} \& \citeyear{Harris13}, Forbes et
al.~\citeyear{Forbes11}, Kartha et al.~\citeyear{Kartha14}).\\

One of the surprising outcomes of those studies is a correlation
between colour and magnitude for GCs of the blue subpopulation
(e.g. Harris et al.~\citeyear{Harris06}, Strader et
al.~\citeyear{Strade06}, Mieske et al.~\citeyear{Mieske06}
\&~\citeyear{Mieske10}, Spitler et al.~\citeyear{Spitle06},
Humphrey~\citeyear{Humphr09}, Cockcroft et al.~\citeyear{Cockcr09},
Harris et al.~\citeyear{Harris09a} \& \citeyear{Harris09b}, Forbes et
al.~\citeyear{Forbes10}, Blom et al.~\citeyear{Blom12}, Usher et
al.~\citeyear{Usher13}). This phenomenon has been nicknamed the
blue tilt, and has also been detected through ground-based
imaging (Forte et al.~\citeyear{Forte07}, Wehner et
al.~\citeyear{Wehner08}, Park~\citeyear{Park12}). The first
  strong hint of the existence of such a colour-magnitude relation of
  GCs was reported in Dirsch et al.~(\citeyear{Dirsch03}), who
  found that the colour distribution of the brightest GCs in NGC 1399
  becomes broad and unimodal. Direct evidence from spectroscopic
  observations was recently reported for M31 GCs
  (Schiavone et al.~\citeyear{Schiav13}).\\

Interpreted as a mass-metallicity relation, the blue tilt is equivalent to
a scaling with mass of $Z \propto M^{0.3-0.7}$, that depends on
environment. Results have so far indicated that the trend sets in already
at masses slightly below a million solar masses, at several $10^5
\msun$. The most common interpretation of this colour-metallicity
relation is self-enrichment. With increasing primordial cluster mass
(and thus increasing potential well depth), a larger fraction of SN II
metal ejecta are kept in the cluster and reprocessed into a second
generation of stars (Strader \& Smith~\citeyear{Strade08}, Bailin \&
Harris~\citeyear{Bailin09}). Furthermore a red tilt, i.e. a colour magnitude
relation for the red GC subpopulation, has been reported in some
cases, but it typically is less significant and weaker in metallicity
space (e.g. Harris et al. 2006, Mieske et al. 2006 \& 2010).\\

It is worth noting that colour changes as a function of GC mass can
also occur as a result of dynamical evolution, via preferential loss of
low-mass stars (Kruijssen \& Lamers~\citeyear{Kruijs08}; Anders et
al.~\citeyear{Anders09}; Kruijssen~\citeyear{Kruijs09}). This makes
lower-mass GCs bluer than their colour for an unaltered stellar mass
function, thus qualitatively creating a trend in the direction of the
blue tilt. This effect was investigated for example in Mieske et
al. (\citeyear{Mieske10}) and particularly in Goudfrooij et
al. (\citeyear{Goudfr14}) who studied how it depends on the shape of
the Initial Mass Function (IMF). However, it is clear that the observed blue tilt can only be
explained in small parts by this effect: the observations indicate a
strengthening of the effect at high cluster masses (Harris et
al.~\citeyear{Harris06}, Mieske et al.~\citeyear{Mieske10}), exactly
opposite to expectations from dynamical evolution: higher mass cluster
are less affected by dynamical evolution (Baumgardt \&
Makino~\citeyear{Baumga03}), and thus colour changes from this effect
will level off at higher cluster masses.\\

An important aspect to keep in mind is that the cluster-to-cluster
scatter of mean metallicity due to local variations in primordial star
forming conditions is significant, and will dominate smooth internal
trends with cluster mass for small samples of globular clusters (see
e.g. Bailin \& Harris~\citeyear{Bailin09}). To filter out underlying
trends like self-enrichment from broadband integrated photometry, one
needs large homogeneous data sets of GCs, typically more than
$\sim$1000 GCs. For example, for the Milky Way no blue tilt is
detectable despite the considerable cluster-to-cluster scatter in mean
[Fe/H] and the comparably small sample size (e.g. Strader \&
Smith~\citeyear{Strade08}). Moreover, the Milky Way GC sample has
  a rather low upper-mass limit ($M_{\omega Cen} = 1.5*10^{6}
  M_\odot$), which is another factor for the
  non-detectability of the blue tilt in our Galaxy.\\

\subsection{Multiple stellar populations in massive Galactic GCs}

Almost parallel in time to the discovery of the blue tilt in
extragalactic GC samples, HST imaging of individual stars in Galactic
GCs also revealed striking new features: there are multiple stellar
populations in almost every intermediate-to-high mass GC that has been
studied in detail (e.g. Bedin et al.~\citeyear{Bedin04}, Piotto et
al. ~\citeyear{Piotto05} \& \citeyear{Piotto12}, Milone et
al.~\citeyear{Milone12a},~\citeyear{Milone12b},~\citeyear{Milone12c}). This
has further enhanced the interest in self-enrichment scenarios for
star clusters. It is typically assumed that self-enrichment from SN
ejecta are required to obtain a broadened stellar metallicity
distribution of red giants as seen for example in $\omega$Cen and M54
(e.g. Willman \& Strader~\citeyear{Willma12}), for which a present-day
onset mass of around a million solar masses is consistent with the
data in the Milky Way. Re-processing of ejecta from evolved stars
  is, in turn, often considered a scenario for creating
multiple populations at constant metallicity (e.g. Ventura et
al.~\citeyear{Ventur01}, d'Antona et al.~\citeyear{dAnton02} \&
\citeyear{dAnton05}, D'Ercole et al.~\citeyear{dErcol08}, Maxwell et
al.~\citeyear{Maxwel14}), which can occur at about one order of
magnitude lower cluster masses than enrichment from SN ejecta
(Baumgardt et al.~\citeyear{Baumga08}).

\subsection{This paper}

In this paper we put special emphasis on how a fit of SN self-enrichment
models to our data constrains the primordial structure and star
formation conditions in massive star clusters.

We analyse the colour-magnitude relation of GCs in the central 100 kpc
of the Hydra and Centaurus galaxy clusters. We use deep U-I imaging
with FORS1@VLT at very good seeing. We analyse a sample of about 2500
GCs in each cluster, down to a limiting magnitude about 0.5 mag
brighter than the GCLF turnover. The usage of the U-band gives us a
very broad wavelength baseline that improves our metallicity
resolution compared to other studies. Furthermore, our homogeneous
data set allows us to perform a robust comparison between the two
investigated environments.

The structure of the paper is as follows. In Sect.~\ref{data} we
discuss the data reduction and selection of GC candidates. In
Sect.~\ref{analysis} we measure the mass-metallicity relations of
GCs in Hydra and Centaurus. In Sect.~\ref{discussion} we constrain a
star cluster self-enrichment model by comparing its prediction to the
measured relation. We focus on the inclusion of previously not
considered physical features into the model, and on the degeneracy
between the input parameters of the model. We also convert the global
mass-metallicity relation into a mass-metallicity {\it spread}
relation, and compare this to Milky Way GCs. We provide Summary and
Conclusions in Sect.~\ref{summary}.


\section{Data}
\label{data}

    \subsection{Observations}

The data analysed for this paper were all taken in service mode with
FORS1@VLT. The I-band photometry was obtained in the course of
programmes 065.N-0459 (Hydra, PI Hilker) and 067.A-0358 (Centaurus, PI
Hilker), with 3000s integration time in I-band per 7x7$'$ FORS
pointing, at good seeing 0.4-0.7$''$ in seven FORS pointings in
  Centaurus and Hydra and one background pointing. Those data were already presented in Mieske et
al. (\citeyear{Mieske03},~\citeyear{Mieske05}) in an SBF analysis of
both clusters. This I-band data set also comes along with V-band data
taken in the same runs and under similar conditions. The V-band data
is not used directly for our blue tilt analysis, but we use the V-I
colour distribution to finetune the U-I colour calibration, see
Sect.~\ref{Uband}. For the present work, these I- (and V-) data were
carefully re-reduced.

The U-band data was obtained in programmes 080.A-0647 \& 082.A-0894 (PI
Hilker) with about 0.8$''$ seeing, in three FORS pointings in
Centaurus, two in Hydra and one background pointing in each
cluster. The on-source integration in U-band was about 2 hours per
pointing. 

\begin{figure*}
\begin{center}
\includegraphics[height = 7.cm] {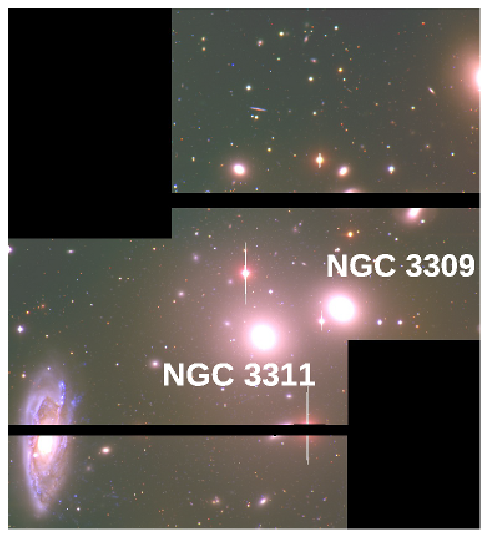} 
\includegraphics[height = 7.cm] {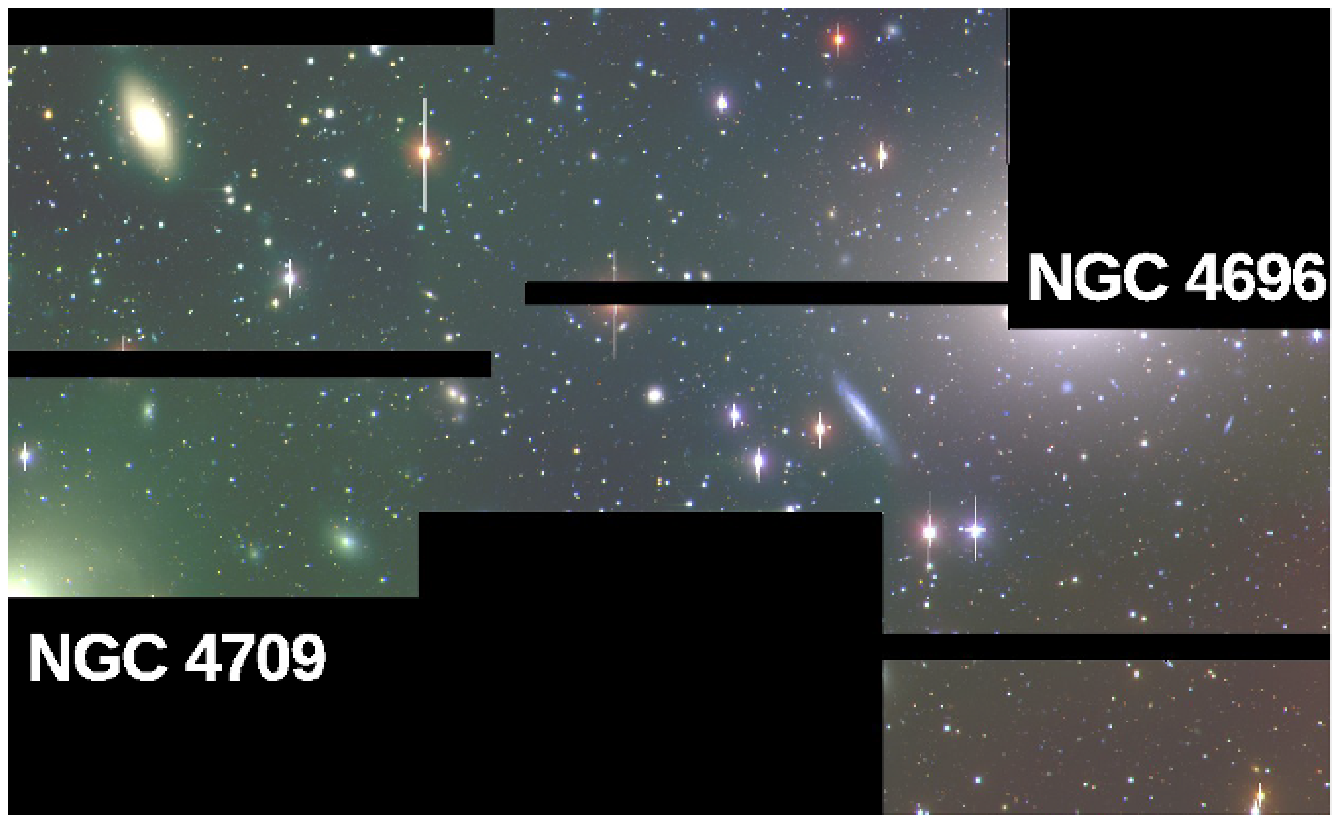}
\end{center}
\caption{\label{img}\small{FORS U,V,I composite-images showing the
    investigated region of the Hydra (left) and Centaurus (right)
    clusters. The two main central cluster galaxies are indicated. For
    Centaurus, the imaging covers the intra-cluster part between NGC
    4696 and NGC 4709, while for Hydra the imaging is centred on NGC
    3311. Projected physical scale is about 100x100 kpc per FORS
    pointing in Hydra, and 90x90 kpc in Centaurus (assuming a NED
    distance of 47.8 Mpc to Hydra and 43.2 Mpc to Centaurus).}}
\end{figure*}

\begin{figure*}
\begin{center}
\includegraphics[width = 17.6cm] {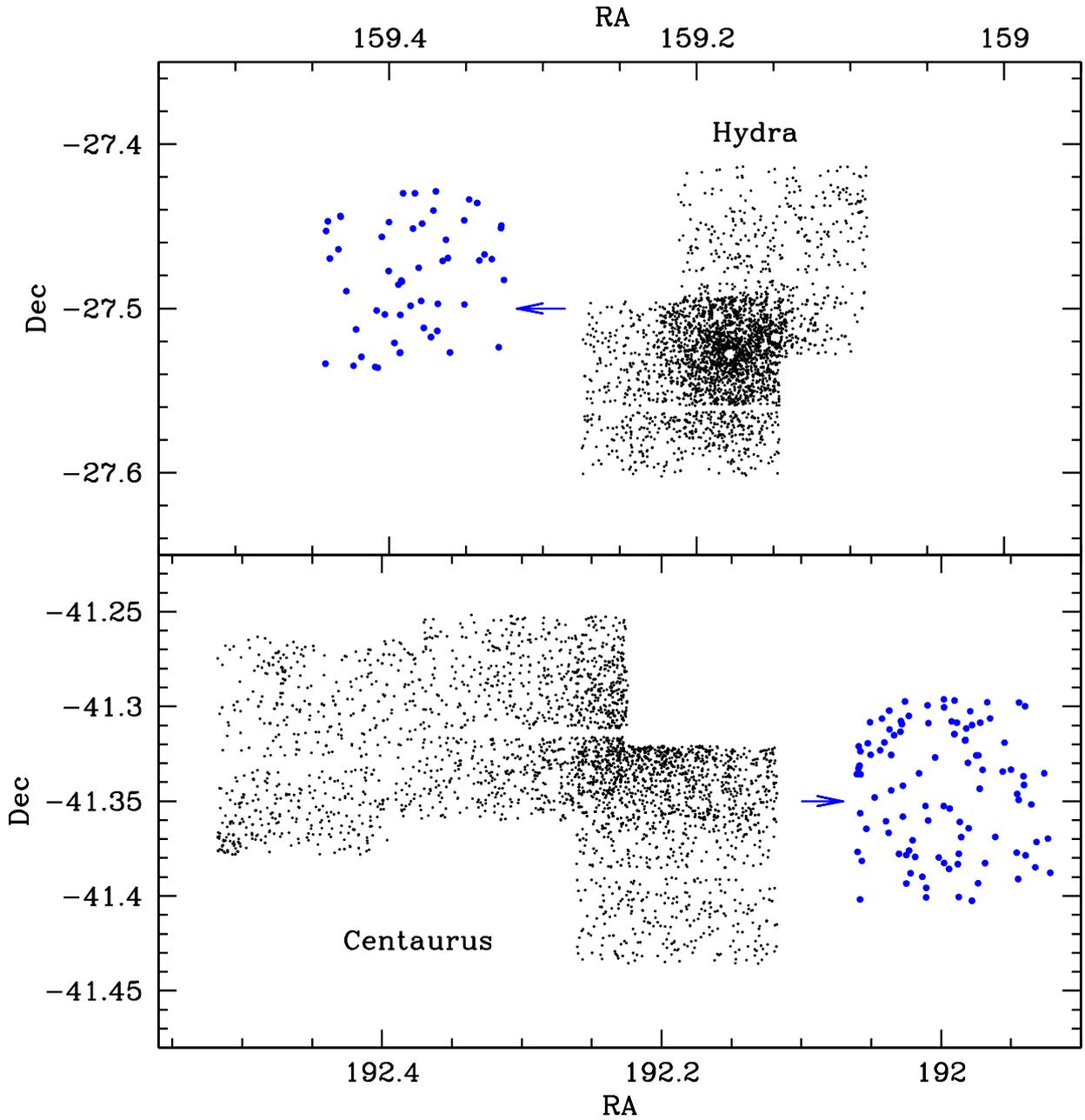}
\end{center}
\caption{\label{map}\small{Projected position of photometrically
    selected GC candidates in Hydra and Centaurus in our U-I FORS1
    imaging data. See text for details on the GC selection. The
    sources detected in the background field with the same photometric
    selection criteria are indicated as (blue) dots for
    comparison. The actual location of the background field in each
    cluster was about 2 degrees away, in the direction indicated by
    the arrow. {\bf Top:} Hydra. {\bf Bottom}: Centaurus. Projected
    physical scale is about 100x100 kpc per FORS pointing in Hydra,
    and 90x90 kpc in Centaurus.}}
\end{figure*}

In Fig.~\ref{img} we show a colour-composite image of the two target
clusters, created from the FORS imaging data in U,V,I. Fig.~\ref{map}
shows a map with the projected positions of GC candidates with U,I
photometry. The background fields are situated at $\alpha
  = 189.78$, $\delta = -41.35$ for Centaurus and $\alpha = 160.9$,
  $\delta = -27.48$ for Hydra, about 2$^{\circ}$ away from the
  respective cluster centres. More details on the reduction and
analysis of the data will be given in a forthcoming paper (Hilker et
al. 2014 in preparation).


    \subsection{Data Reduction}

The raw data were processed with IRAF\footnote{IRAF is distributed by
  the National Optical Astronomy Observatories, which are operated by
  the Association of Universities for Research in Astronomy, Inc.,
  under cooperative agreement with the National Science
  Foundation.}. We performed standard bias subtraction, flat fielding
and bad pixel removal. We then registered and combined the single
integrations. The next step was to model and subtract the extended
light from bright galaxies, to enable clean photometry of point source
close to the galaxy centers. We did so using a combination of IRAF's
\emph{rmedian} task, and galaxy light modelling with IRAF's
\emph{ellipse} task. Diffuse galaxy light is thus removed apart from
their very central regions. The PSF photometry of point sources was
performed with the standard IRAF routine \emph{psf} within the
\emph{daophot} package.

Photometric calibration was done using regular standard star
observations within ESO's calibration plan. We corrected relative
offsets between different pointings (not all nights were photometric)
by using overlap regions of adjacent fields, and using the central
pointing in each cluster as reference. These relative calibrations
worked very well for the I-band with residual inaccuracies of the
level of 0.02-0.03 mag. However, the U-band data could not be
calibrated down to this accuracy, given the low number of standard
stars available in U, lower flux levels, and smaller overlap between
adjacent fields. Therefore we applied an additional, relative U-I
colour scale calibration between Hydra and Centaurus as described in
the following.

\subsubsection{Relative calibration of U-I colour scale}
\label{Uband}

\begin{figure}
\begin{center}
\includegraphics[width = 4.3cm] {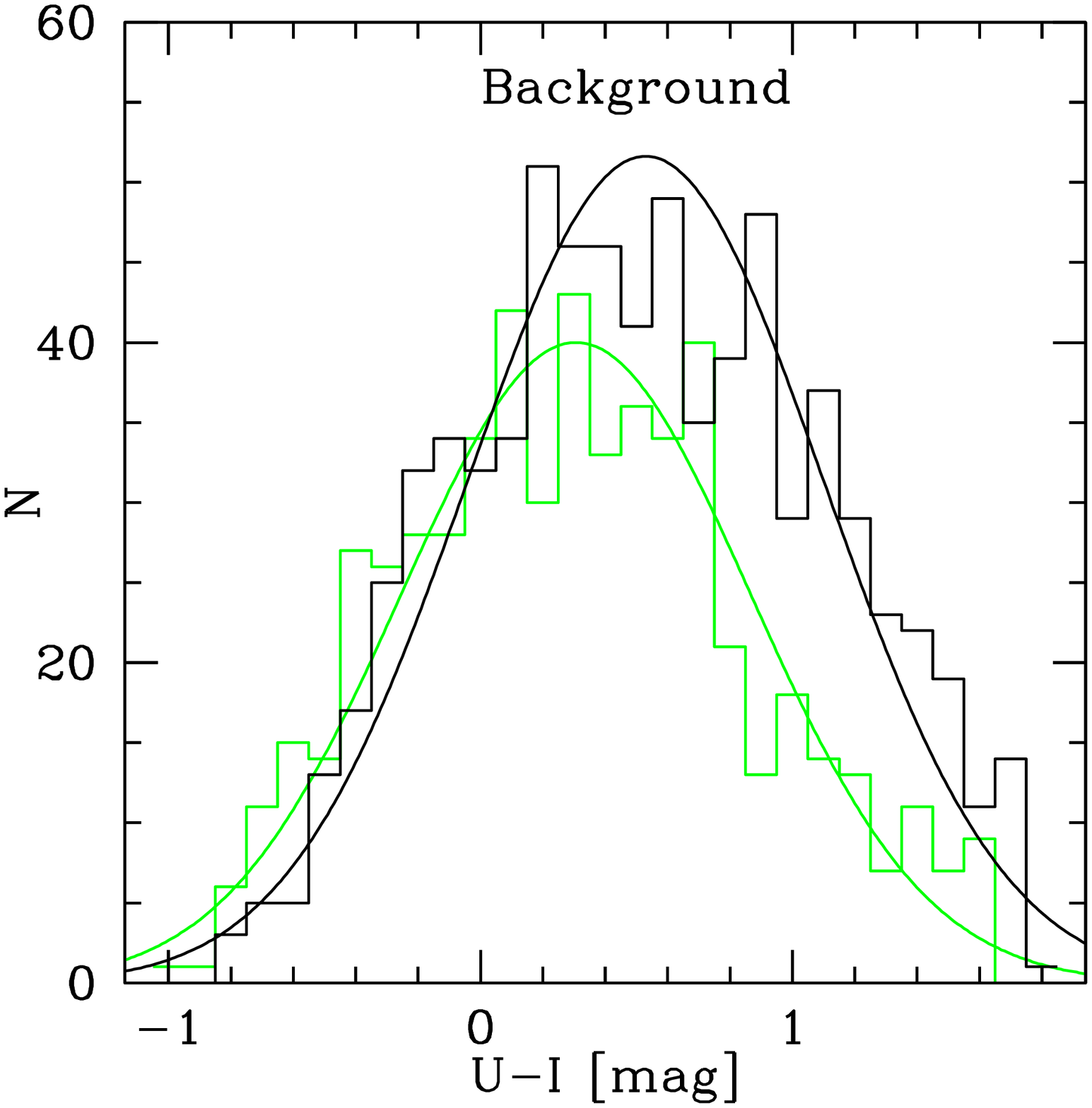}
\includegraphics[width = 4.3cm] {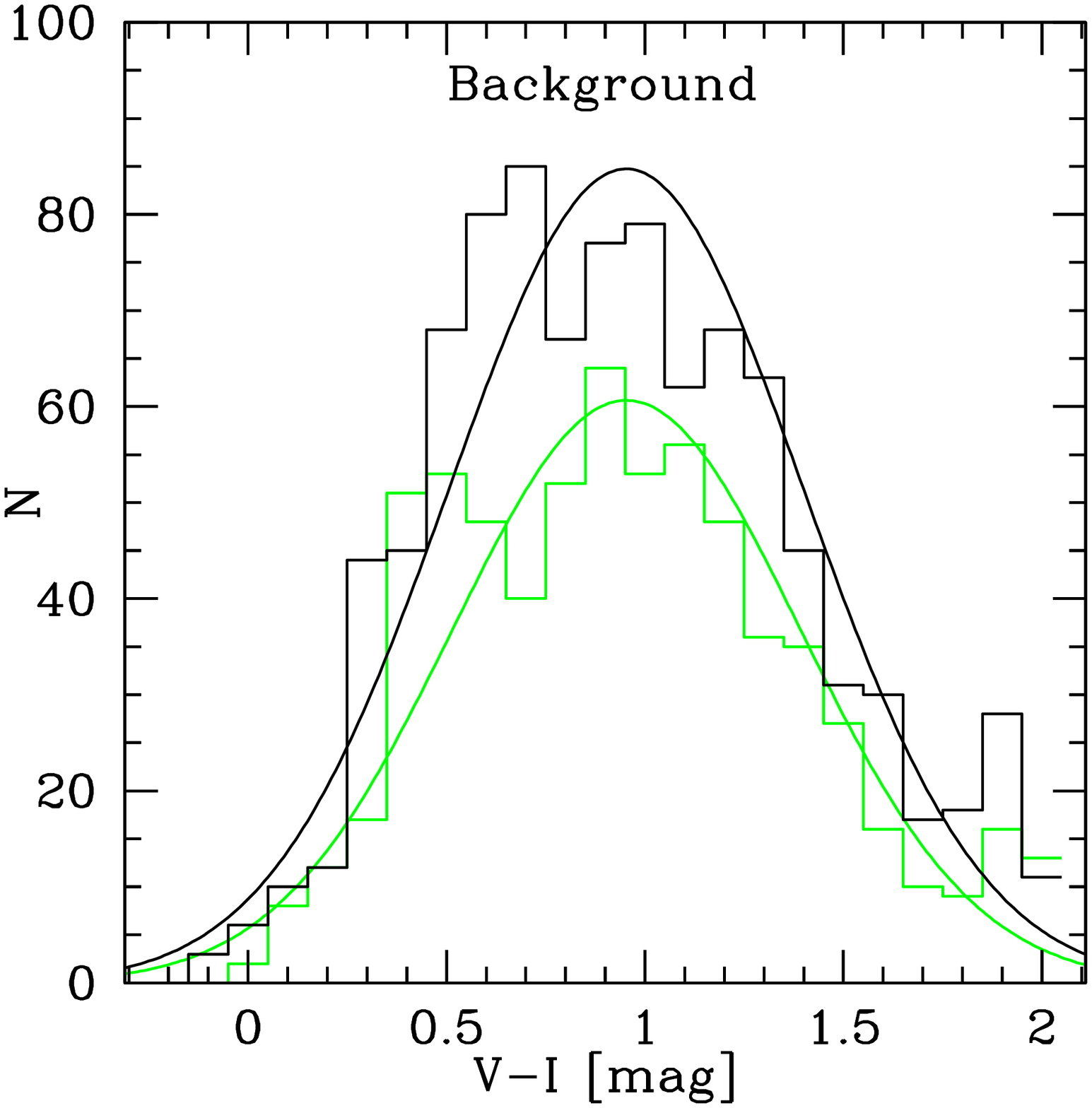}
\includegraphics[width = 4.3cm] {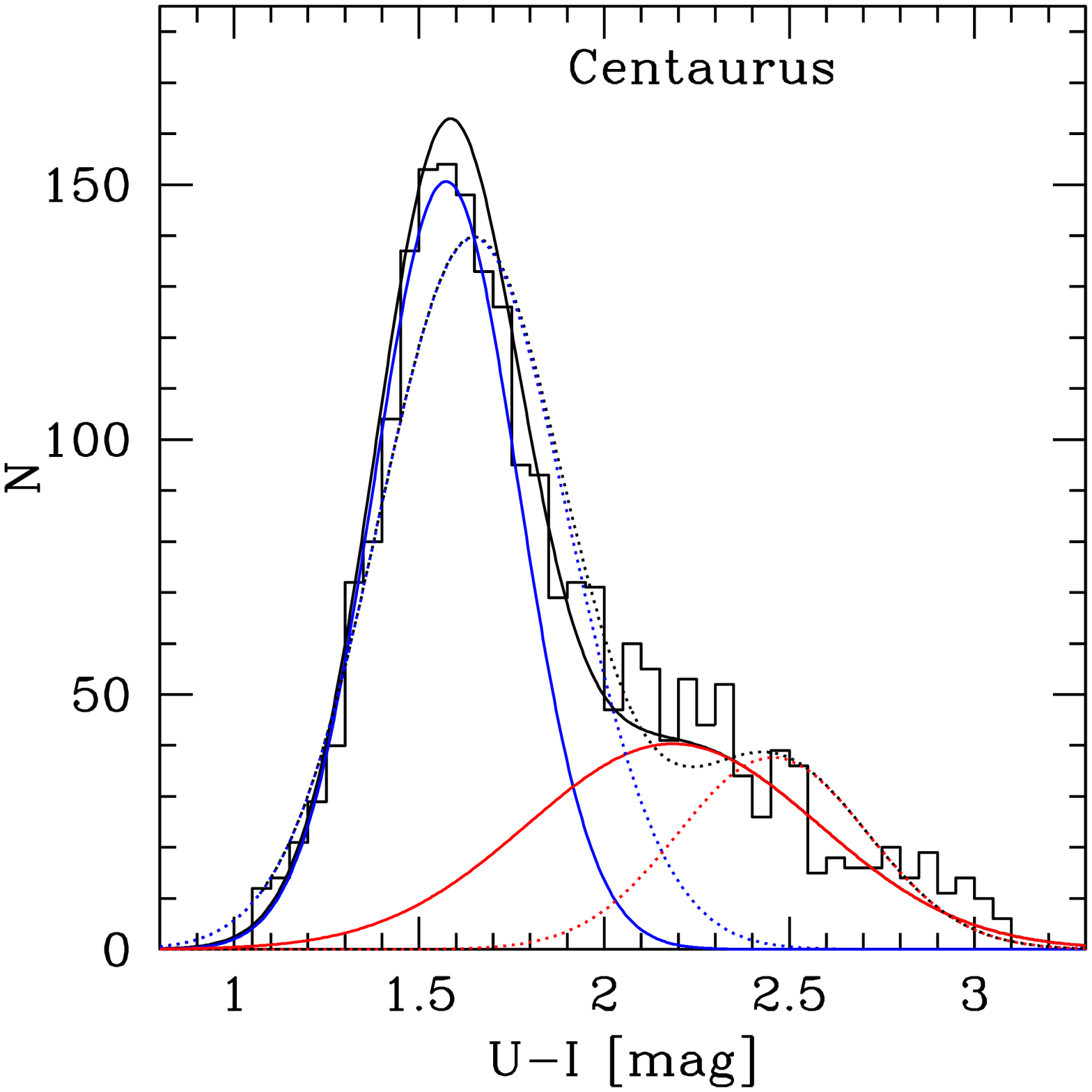}
\includegraphics[width = 4.3cm] {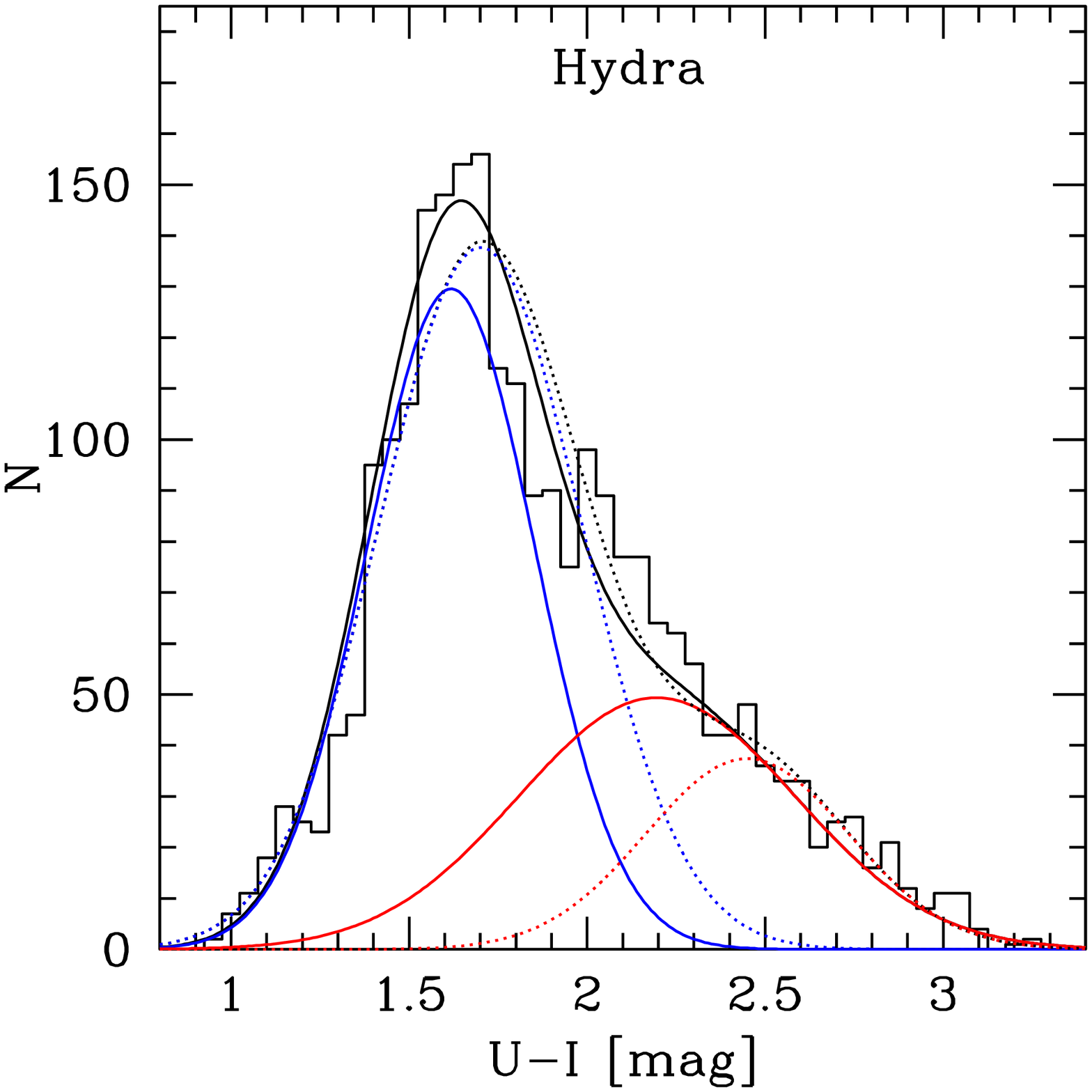}
\end{center}
\caption{\label{bgcol}\small{{\bf Top:} Colour distribution of the
    background fields: Centaurus is in black and Hydra in green. {\bf
      Left panel:} U-I. {\bf Right panel:} V-I. The lines correspond
    to single Gaussian peaks fitted with KMM.} Because of the shift in U-I
  between both clusters, symmetric shifts of $\pm$0.11 mag are applied
  to the U-band zeropoints for Hydra and Centaurus, see text for
  details. {\bf Bottom:} Finally adopted U-I colour distribution of the
  full Hydra (left) and Centaurus (right) cluster GC samples
  (restricted to the later used fitting range $I<24$ mag). The coloured
  lines represent the Gaussian fits by \emph{KMM} to the blue and red
  peak. The black lines are the sum of the two. The dotted lines
  represent the results using the homoscedastic mode; the solid lines
  correspond to the heteroscedactic mode.}
\end{figure}

  The top panels of Fig.~\ref{bgcol} show the U-I (left) and V-I
  (right) colour distribution of sources detected in the background
  fields for Hydra and Centaurus. While the V-I distribution of both
  samples peaks at the same colour within the error bars, there is a
  shift of about 0.2 mag in U-I between Hydra and
  Centaurus. Using KMM (Ashman et al.~\citeyear{Ashman94}), the precise value of the
  shift is determined to be 0.22 mag. We interpret this colour shift as
  a residual systematic uncertainty of the U-band photometric
  calibration in Hydra and Centaurus, as noted in the previous subsection. In
  order to compare the two samples on the same absolute colour scale,
  we add to the U-band an offset of -0.11 mag to Centaurus sources and
  +0.11 mag to Hydra in the following. The bottom panels of
  Fig.~\ref{bgcol}, which show the Centaurus and Hydra GC colour
  distribution in U-I, include those offsets.

    \subsection{Selection of Globular Clusters}

     	The first step was to remove resolved sources by a combined
        cut-off in the IRAF PSF parameters $\chi$ and
        \emph{Sharpness}. This reduced the contamination by a factor
        of two.

\begin{figure*}
\begin{center}
\includegraphics[width = 8.6cm] {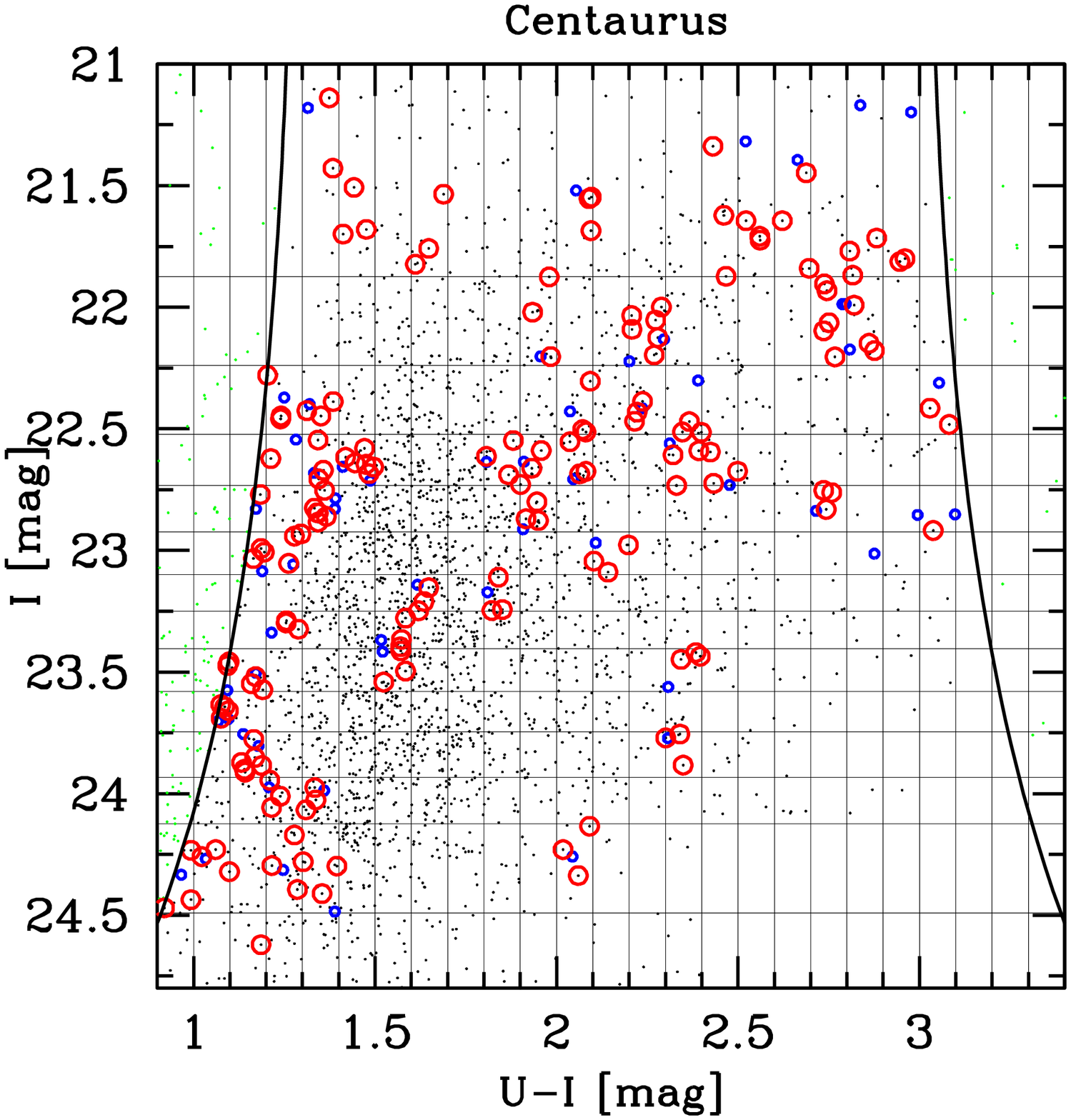}
\includegraphics[width = 8.6cm] {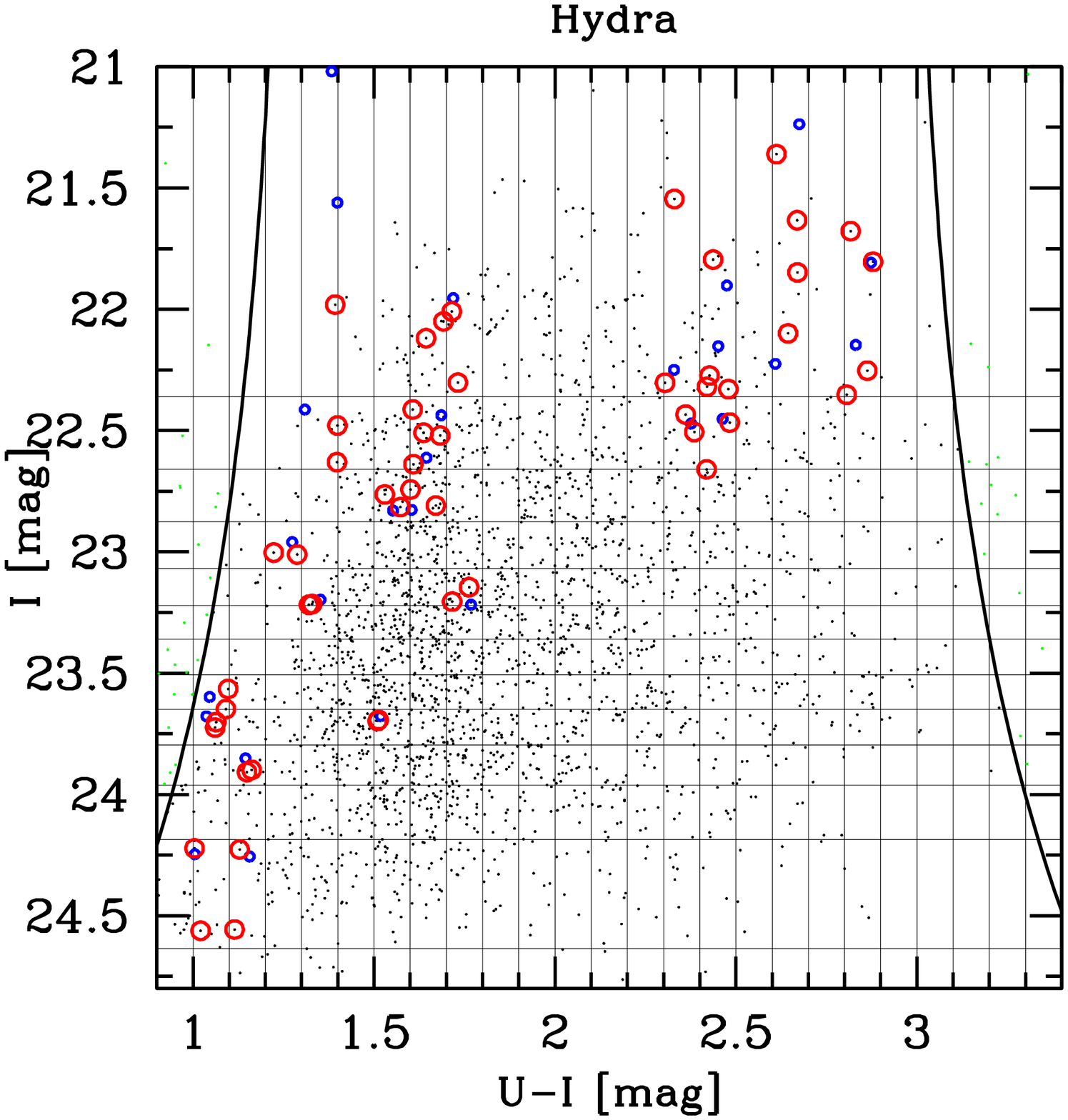}
\end{center}
\caption{\small{Colour-Magnitude diagram in UI of all unresolved
    sources detected in both galaxy clusters. The black curved lines
    represent the colour cut-off applied to disregard sources either
    too blue or too red to be considered old GCs. The horizontal lines
    define rows of 200 GCs, and the vertical lines define columns of
    0.1 mag. Within the cells defined by the horizontal and vertical
    lines, a random background subtraction is performed: The blue
    circles indicate colours and magnitudes of the background
    sources. Red circles indicate sources that were removed based on
    the background source distribution, within the indicated
    cells. We note that the very brighest magnitude regions $I<21 mag$
    are excluded from the plot to focus on the main body of GCs.}}
\label{compare_CMD_UI}
\end{figure*}

     	The second step was to introduce a realistic colour cut-off
        condition that corresponds to the extreme U-I colours expected
        for GCs. Such values were taken from the metallicity
        distribution of Milky Way GCs obtained by Harris
          (\citeyear{Harris96}, 2010 edition), that is
        $\big{[}\frac{Fe}{H}\big{]} = -2.4 $ and
        $\big{[}\frac{Fe}{H}\big{]} = 0.3 $, converted into a U-I
        colour range of 1.3 to 3.0 mag using TERAMO models (Raimondo et
        al.~\citeyear{Raimon05}) evaluated at 11-13 Gyr age
        (Figs.~\ref{compare_CMD_UI} and ~\ref{teramo}). We broadened
        these borders by the maximum measurement error as a function
        of luminosity, as indicated by the solid lines in
        Fig.~\ref{compare_CMD_UI}.

     		To remove the remaining contamination by unresolved
                foreground stars and background galaxies in the colour
                range expected for GCs, a comparison between the
                science field and the background field was done.  For
                this we divided the colour-magnitude plane in cells
                with rows of 200 GCs and 0.1 mag colour width, see
                Fig.~\ref{compare_CMD_UI}. Sources in Hydra and
                Centaurus were then randomly deleted in each cell
                according to the number of background points in the
                cell, and the respective ratio between the area
                coverage of cluster and background sample: 3:1 in
                Centaurus and 2:1 for Hydra. 

After background subtraction, the final GC candidate sample comprised
2590 sources for Centaurus, and 2365 sources for Hydra.
                
\section{Data Analysis}
\label{analysis}
\subsection{Fitting Procedure}

To measure the colour of the blue and red peak as a function of GC
luminosity, we applied a hybrid approach using both KMM and colour
distribution medians (see also Mieske et
al.~\citeyear{Mieske06}). This is outlined in the following.

In the bottom panel of Fig.~\ref{bgcol} we show the U-I colour
histogram of the GC candidates in Hydra and Centaurus, limited to
$I<24$ mag. The distribution is not obviously bimodal, but does have a
large tail towards red colours that cannot be fitted by a single
Gaussian. In that figure we indicate the fitted colour peaks using
KMM, both in heteroscedastic and homoscedastic mode. It is clear that
the output of both modes differ, indicating that the width of the blue
and red peak are not identical. Therefore, a heteroscedastic fit is
more appropriate to our sample than a homoscedastic fit. 

At the same time, we found that such heteroscedastic fits increase the
scatter between adjacent luminosity bins, i.e. when subdividing the
sample into a number of smaller subsamples. Our aim in the fitting is
to represent the heteroscedastic nature of our sample, but at the same
time provide a stable peak determination as a function of luminosity.
Therefore we measured in each luminosity bin of 200 GCs the median
colour blue- and red-wards of a constant
  dividing line between blue and red sequences (see also
Fig. 5 of Mieske et al.~\citeyear{Mieske06}), itself determined from a
heteroscedastic KMM fit to the full sample. We adopt such a constant
  dividing line - as opposed to a
limit that varies as a function of luminosity - to ensure that a
distribution without any colour-magnitude relation would be correctly
recovered.  This limiting colour is U-I = 1.93 mag for Centaurus and
  1.99 mag for Hydra (Fig.~\ref{bgcol}). It corresponds to about
  $[Fe/H]=-0.9$ dex, within $\sim$0.1 dex of the limit found in the
  Milky Way GC system (Harris et al.~\citeyear{Harris96}, see
  e.g. Fig.2 of Strader \& Smith~(\citeyear{Strade08}). Luminosities
  fainter than I=24 mag are not included for the fit because of a lack
  of faint red clusters (incompleteness) that would introduce a bias
  in our analysis.

\begin{figure*}
\begin{center}
\includegraphics[width = 8.6cm] {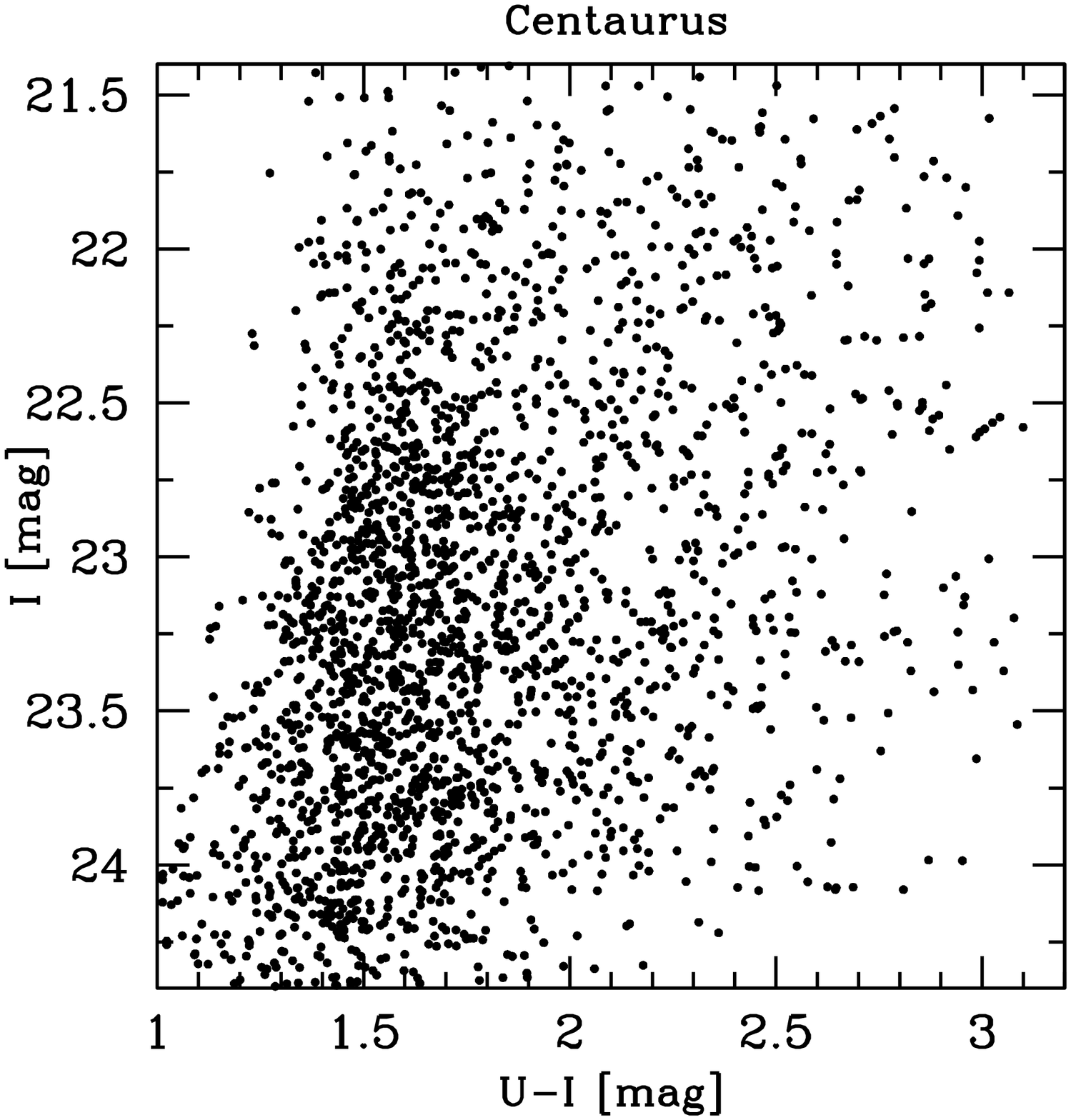}
\includegraphics[width = 8.6cm] {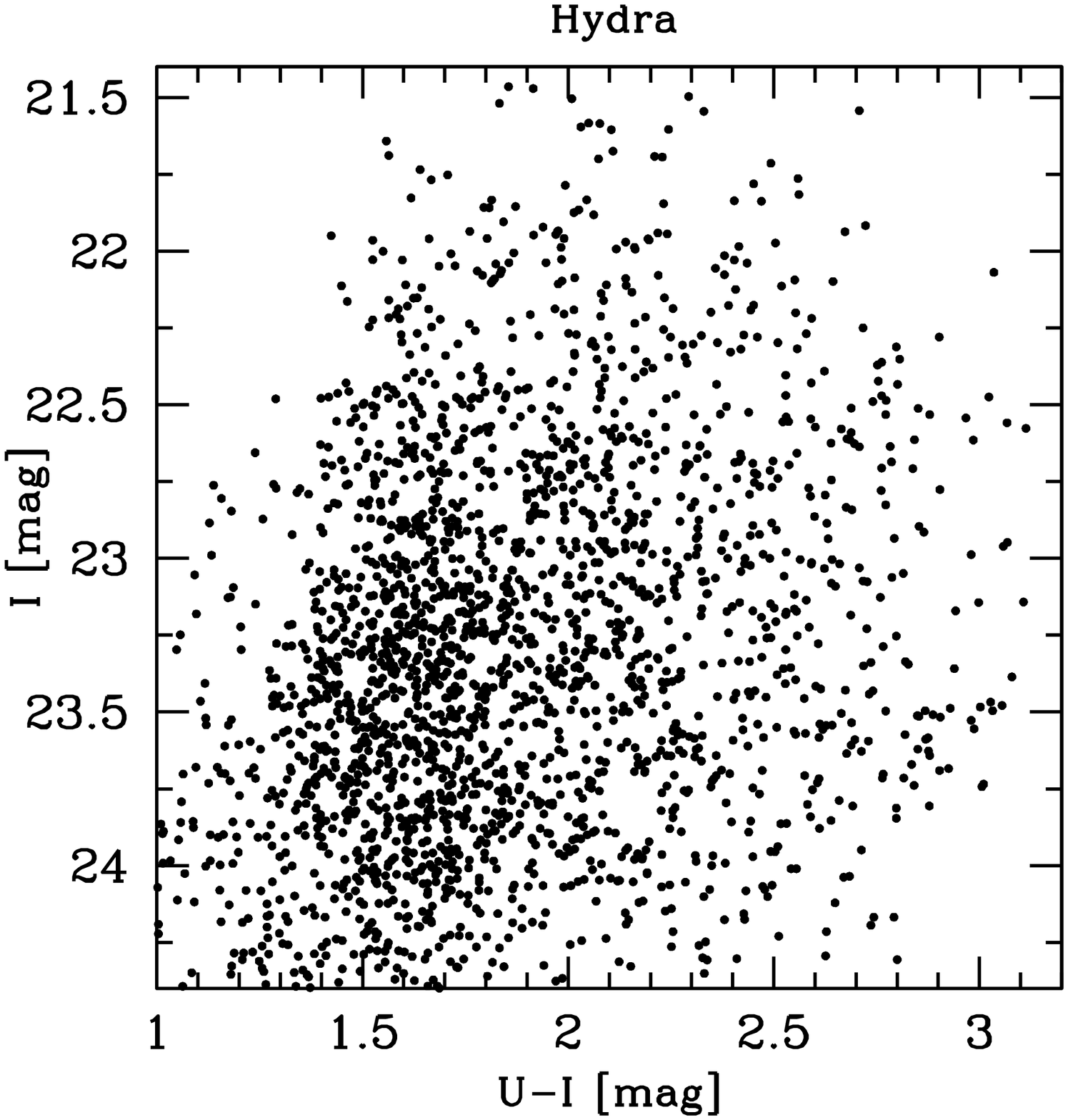}
\includegraphics[width = 8.6cm] {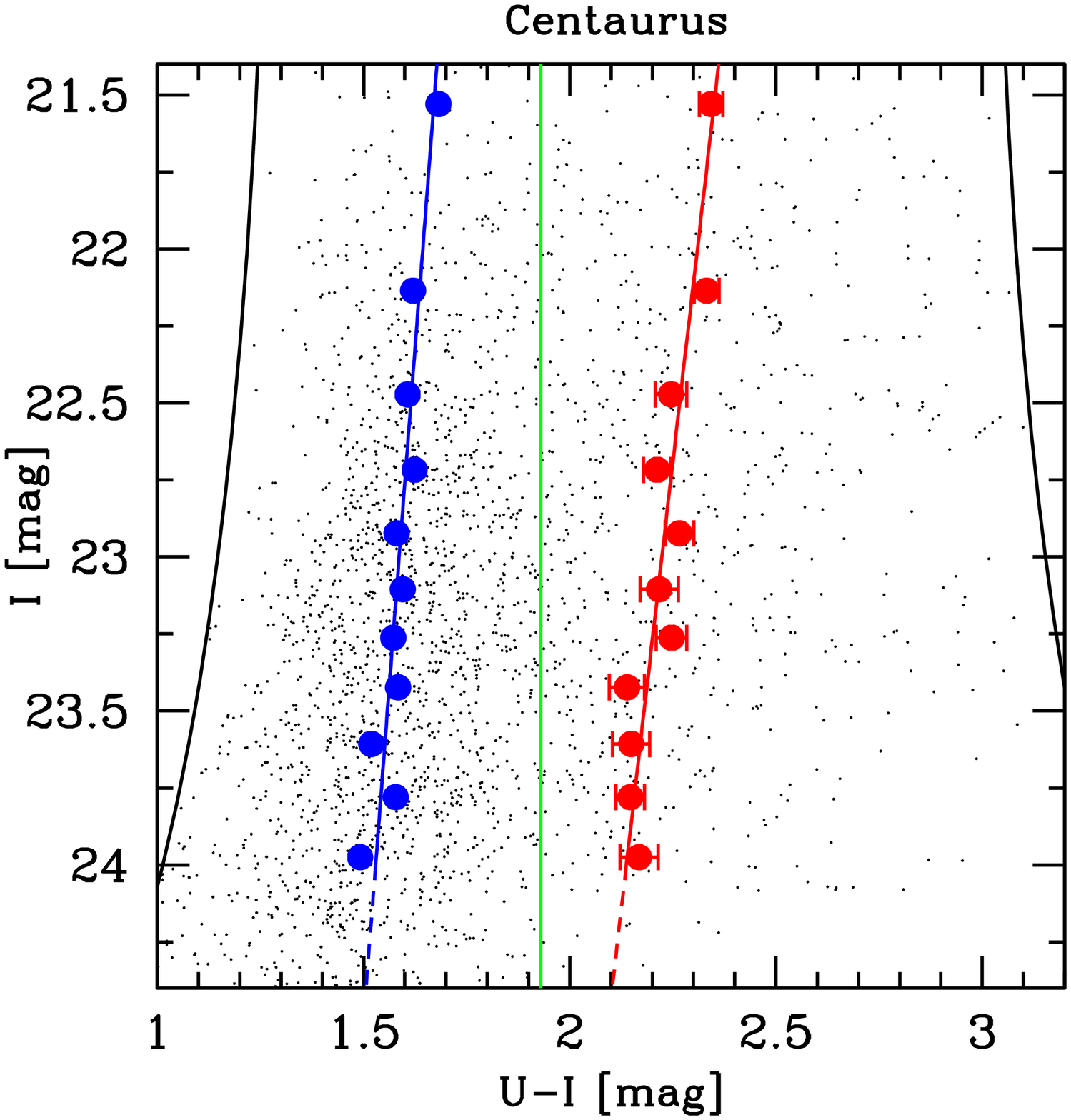}
\includegraphics[width = 8.6cm] {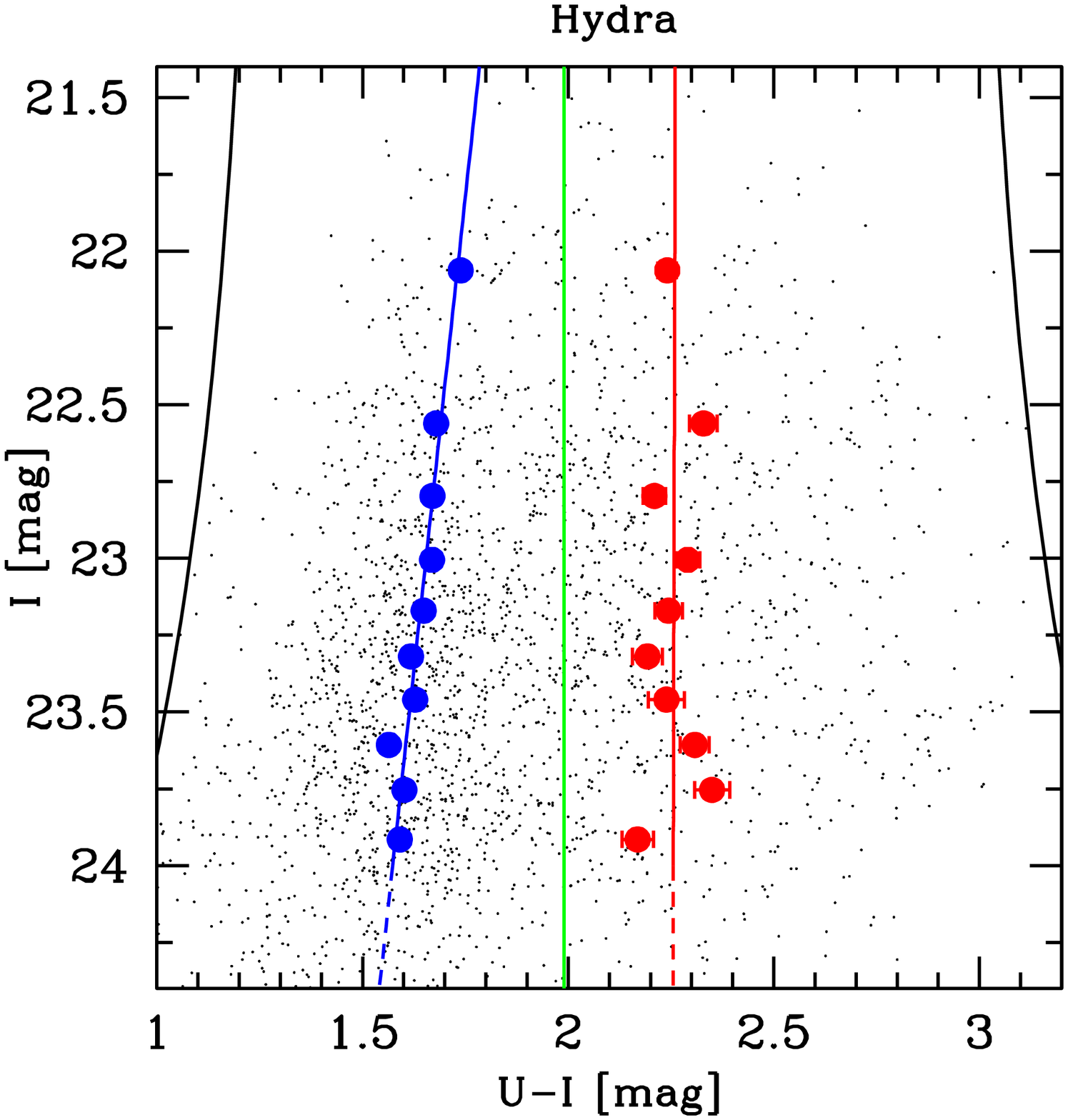}
\end{center}

\caption{\small{Colour-Magnitude Diagrams of the final GC samples for
    Centaurus (left) and Hydra (right). The individual data
      points in the top and bottom plot are the same. In the bottom
      plots additional filled circles in blue and red indicate the
    median colours between the black limiting curves at extreme
    colours and the (green) fixed dividing line between the blue and
    red peak. The blue and red lines indicate least squares fits
      to the median colours as a function of magnitude.}}
\label{compare_CMD}
\end{figure*}

In Fig.~\ref{compare_CMD} we show the resulting blue and red peak
colours as a function of GC luminosity for the Hydra and Centaurus
samples. We note that this distribution is not 100\% deterministic
because of the background source subtraction that is performed
randomly within colour-magnitude cells as desribed in
Fig.~\ref{compare_CMD_UI}. To take this into account for the fitting
of the slope between colour and magnitude, we perform this fitting
based on 100 different random background subtractions. To each
individual run, we apply an unweighted least squares fit (each
magnitude bin comprises the same number of data points) of a linear
relation to these data points, whose slope we denote as $\gamma =
d(U-I) / d(I)$.

We indicate the resulting average slopes and their error bars in
Table~\ref{table_slope}, and also indicate the (small) additional
statistical error arising from the random background subtraction.  We
obtain slopes for the blue peak of \(\gamma = -0.057 \pm 0.011 \) for
Centaurus and \(\gamma = -0.081 \pm 0.012 \) for Hydra, thus finding a
highly significant blue tilt in both clusters. A red tilt is also
detected for Centaurus, with \(\gamma = -0.086 \pm 0.015 \). The error
bars are derived from the scatter of the data points with respect to
the fitted relation.

\begin{figure}[h!]
\begin{center}
\includegraphics[width = 8.6cm] {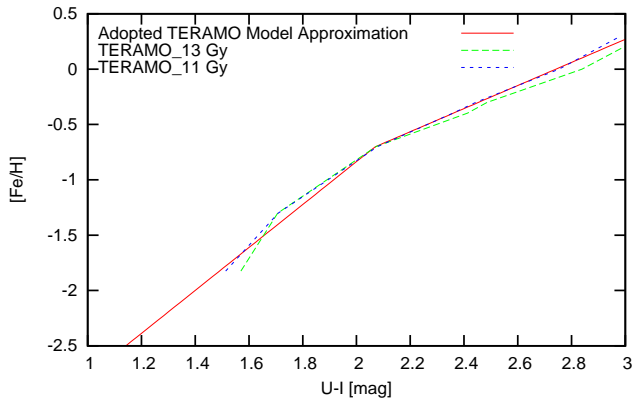}
\end{center}
\caption{\small{Metallicity [Fe/H] as a function of U-I colour
    according to TERAMO Models (Raimondo et al.~\citeyear{Raimon05})
    for 11 and 13 Gyr old single stellar populations. The solid red
    line indicates the adopted piecewise linear approximation that we
    adopt for the transformation from U-I to [Fe/H] in this paper, see
    equations~\ref{colmetblue} and \ref{colmetred}.}}
\label{teramo}
\end{figure}

\begin{figure}[h!]
\begin{center}
\includegraphics[width = 8.6cm] {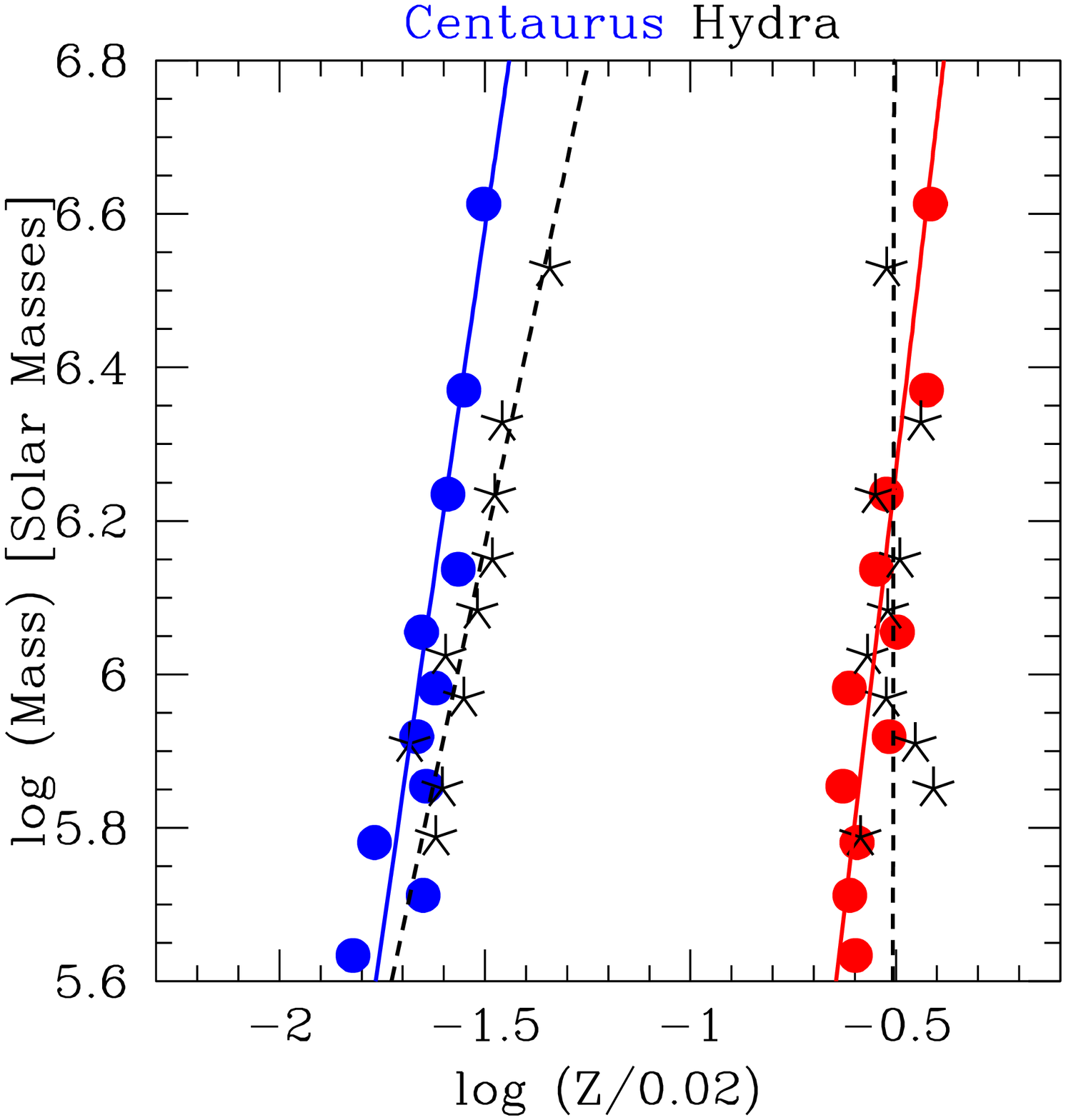}
\end{center}
\caption{\label{mmr}\small{Mass-metallicity relations of GCs in
    Centaurus and Hydra, as converted from the UI colour magnitude
    relations shown in Fig.~\ref{compare_CMD}. Centaurus is shown in
    blue/red filled circles, Hydra in black asterisks. Lines indicate
    unweighted linear fits to the data.}}
\end{figure}

        \subsection{Mass-Metallicity Relations}

   In the previous subsection we find a clear blue tilt for the Hydra
   and Centaurus cluster. To investigate this in more detail it is
   necessary to convert the colour-magnitude relation into a more
   physical mass-metallicity relation.

   To this end, we convert I-band magnitude into mass using:

\begin{itemize}
\item A colour-independent mass-to-light ratio of $\Gamma =2.2$, based
        on the models of Maraston (\citeyear{Marast05}) and Bruzual \& Charlot
        (\citeyear{Bruzua03}).
 \item A NED distance of 47.8 Mpc to Hydra and 43.2 Mpc to
        Centaurus. 
\item An absolute magnitude $M_I=4.13$ mag of the
        Sun. The resulting mass range of GCs is $\sim 5 \times 10^5$
        to 5$\times 10^6$ solar masses.
\end{itemize}

Colour U-I is converted into metallicity using a piecewise linear
function representing the TERAMO Models (Raimondo et
al.~\cite{Raimon05}) for an assumed age of 11-13 Gyr, see
Fig.~\ref{teramo}. We adopt two lines of different slopes that
intersect at U-I=2.07, which is accurate to within 0.1 dex compared to
the model predictions. The conversion from U-I to [Fe/H] thus is
\begin{equation}
[Fe/H]=-0.702 + 1.94 \times [(U-I)-2.0688]
\label{colmetblue}
\end{equation}
for $(U-I)<= 2.0688$  $([Fe/H] \leq -0.702)$. And,
\begin{equation}
[Fe/H]=-0.702 + 1.04 \times [(U-I)-2.0688]
\label{colmetred}
\end{equation}
for $(U-I)>2.0688$ $([Fe/H]>-0.702)$.\\

We note that a polynomial fit is problematic for the extrapolation to
colours bluer than the model limits at U-I$\lesssim 1.55$ mag ($[Fe/H]
\lesssim -1.8$ dex), which is why we prefered a piecewise linear
approximation of the models.

\begin{table*}
\caption{\label{table_slope}\small{Value of the blue tilt slope
    $\gamma = d(U-I)/dI$ and of the Mass-Metallicity relation exponent
    \(\alpha\) obtained for Centaurus and Hydra with our data for the
    range $I<24.0$ mag, compared to results of previous studies on
    these clusters and other environments. In brackets we also
    indicate the errors arising from the randomised background
    subtraction in magnitude-colour cells.}}
\begin{center}
\begin{tabular}{c|c|c|c|c}
Sample  & \(\gamma\) blue &  \(\alpha\) blue ( \( Z \propto M^\alpha \)) &  \(\gamma \) red &  \(\alpha\) red  ( \( Z \propto M^\alpha \)) \\\hline

Centaurus    &  -0.057  $\pm$ 0.011 [$\pm$ 0.003] &   0.27 $\pm$ 0.05   & $-$0.086  $\pm$ 0.015 [$\pm$ 0.002]  & 0.22 $\pm$ 0.04 \\

Hydra        &  -0.081 $\pm$ 0.012 [$\pm$ 0.005] &   0.40  $\pm$ 0.06  & $-$0.002 $\pm$ 0.034 [$\pm$ 0.004] & 0.01 $\pm$ 0.09  \\

Hydra (Wehner et al. 2008) & & 0.60 $\pm$ 0.20 & & \\

 BCGs (Harris et al.\citeyear{Harris09a}) & &  0.48 $\pm$ 0.12 & & $-$0.11 $\pm$ 0.06 \\

 M104 (Harris et al.\citeyear{Harris10}) & & 0.29 $\pm$ 0.04 & & \\

FCS (Mieske et al. 2010)      &     & 0.38 $\pm$ 0.08    &    & 0.03 $\pm$ 0.05 \\

VCS (Mieske et al. 2010)      &     & 0.62 $\pm$ 0.08    &    & 0.12 $\pm$ 0.05 \\  
\end{tabular}
\end{center}
\end{table*}

The resulting data points in mass-metallicity space are shown in
Fig.~\ref{mmr} for both Hydra and Centaurus along with a linear fit to
the data. The slopes of the linear fits are listed in
Table~\ref{table_slope} in terms of the exponent $\alpha$ in $Z
\propto M^\alpha$. In particular we find $Z \propto M^ {0.27 \pm
  \bf 0.05}$ for GCs in Centaurus and $Z \propto M^ {0.40 \pm 0.06}$ for
GCs in Hydra along the blue sequence.


\begin{figure*}
\begin{center}
\includegraphics[width = 8.4cm] {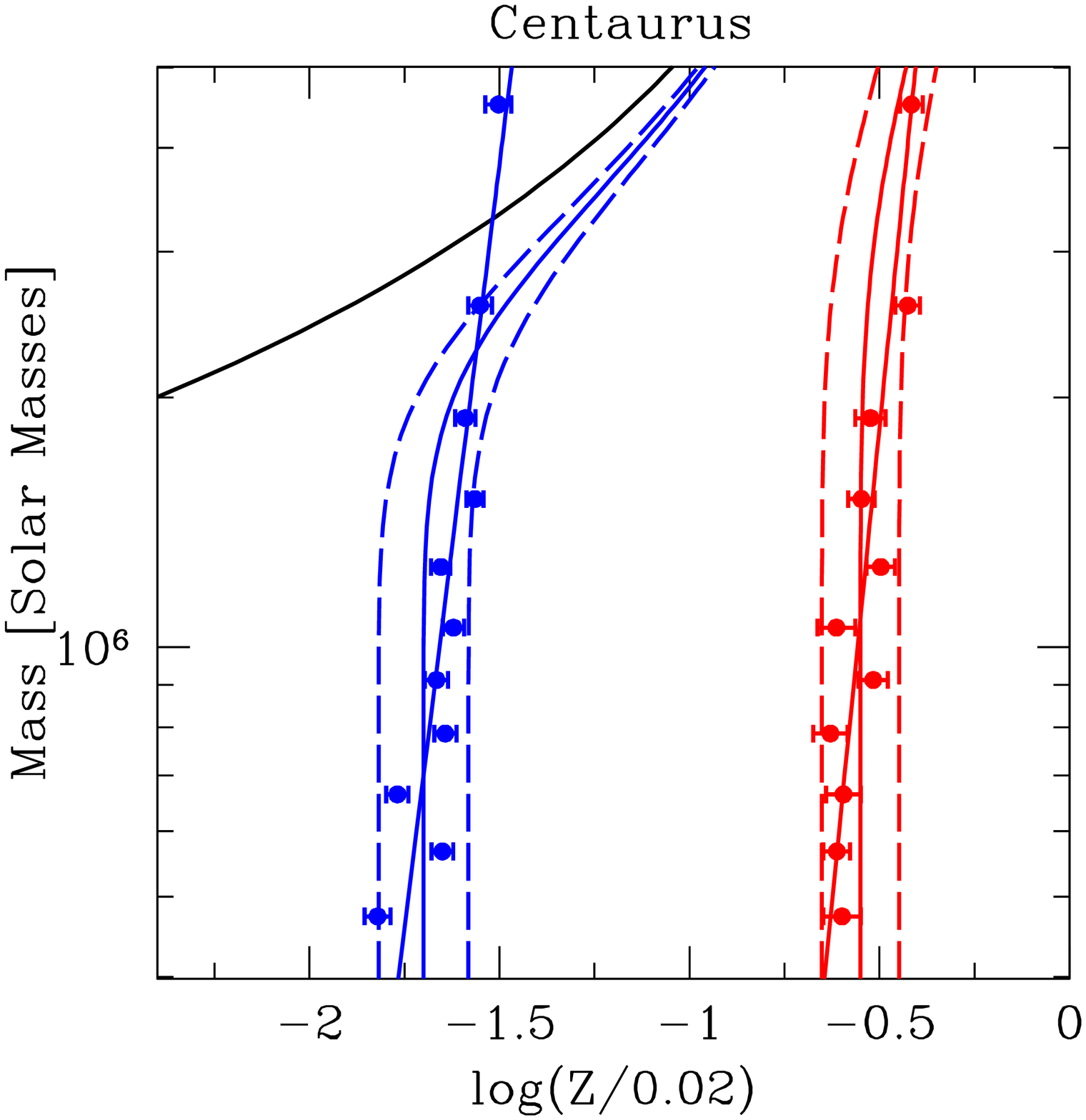}
\includegraphics[width = 8.4cm] {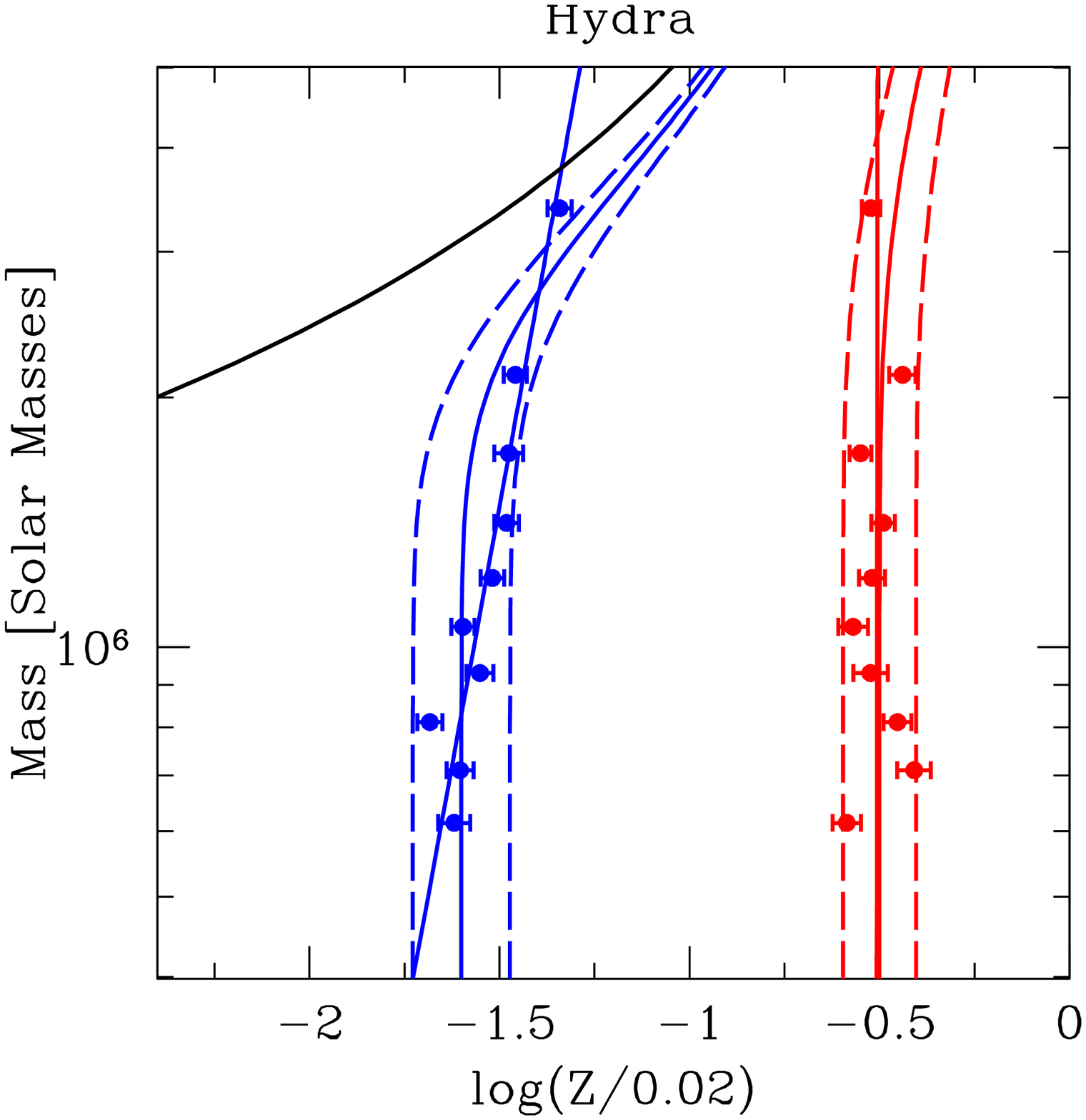}
\end{center}
\caption{\small{Comparison between the measured mass-metallicity data
    points of GCs in Hydra/Centaurus, and the default version of the
    self-enrichment model by Bailin \& Harris. The Y axis is the
    current mass of the GCs. The black curve shows the self-enrichment
    model with $\beta = 2$ and \(f_\star\) =0.3. The blue and red solid
    curves are the result of the combination between self-enrichment
    and pre-enrichment. The dashed lines represent the scatter in
    colour of the points used in Fig.~\ref{compare_CMD}, converted into
    metallicity. The straight lines are the linear fits from
    Fig.~\ref{mmr}.}}
\label{default}
\end{figure*}

\section{Discussion}
\label{discussion}
In this Section we compare our findings to predictions of the
self-enrichment model of Bailin \& Harris (\citeyear{Bailin09}). This
includes a detailed discussion of how our findings can be used to
constrain GC mass and radius evolution and primordial star formation
efficiencies in the context of that model.

     \subsection{The default model parameters of Bailin \& Harris (2009)}

     Both in Strader \& Smith (\citeyear{Strade08}) and Bailin \&
     Harris (\citeyear{Bailin09}) self-enrichment scenarii of star
     clusters are discussed with a focus on the blue tilt. The basic
     idea of self-enrichment is simple. Self-enrichment (by SNII
     ejecta) starts to become efficient when the gravitational binding
     energy ($\propto$mass) of and gas ram pressure within the
     primordial cluster becomes comparable to the kinetic energy
     output of SNII. In the following detailed discussion we focus on
     the more recent model of Bailin \& Harris.

To calculate the resulting metallicity in a globular cluster as a
consequence of self-enrichment, Bailin \& Harris used as factor of
parametrisation the metal-retention fraction \(f_Z\), which is
obtained from the competition between SNII kinetic energy output
gravitational potential, (equation 28 of their paper):

\Large
\begin{equation}
f_Z =  e^{-\frac{f_\star E_{SN} r_t}{10^2 M_\odot GM_{C}}}
\label{fzsimple}
\end{equation}
\normalsize

This expression is valid for a density profile of an isothermal sphere
( \(\rho \propto r^{-2}\)). $f_\star$ denotes the star formation
efficiency, i.e. the fraction of primordial gas that is converted to
stars. \(E_{SN} = 10^{51}\) ergs is the typical energy released by one
SN II, $M_C$ is the mass of the proto-cluster cloud, and $r_t$ is
  the truncation radius, which is identified with the half-light
  radius $r_h$ in the following (see also Mieske et
  al.~\citeyear{Mieske10}). The factor $10^2$ in the denominator of
the exponent comes from the fact that (for a Salpeter/Kroupa type
high-mass IMF) about one SN is formed per 100 solar masses formed in
stars (Fig.3 of their paper). For \(\beta \neq 2\), one obtains
\(f_Z\) (equation 36 of their paper):

\begin{equation}
f_Z = ( 1 - \frac{f_\star^2E_{SN}r_h(2-\beta)}{10^2 M_\odot GM_{GC}}) ^{\frac{3-\beta}{2-\beta}}
\label{f_z}
\end{equation}

Both $f_z$ and $f_*$ are then used as scaling factor in the following
basic equation describing the final cluster metallicity $Z_C$ obtained
by including self-enrichment (equation 7 of Bailin \& Harris):

\begin{equation}
log_{10}(\frac{Z_C}{Z_{\odot}} ) = 0.38 + log_{10}( f_\star f_Z) + log_{10}(\frac{Z_{pre}}{Z_{\odot}} )
\label{fracZ}
\end{equation}  

where 0.38 is adopted from previous studies (Woosley et
al.~\citeyear{Woosle95} and Nomoto et
al.~\citeyear{Nomoto97}). $Z_{pre}$ is the pre-enrichment of the
globular cluster, i.e. the average metallicity of its first generation
of stars. In their default model, Bailin \& Harris assume
$f_\star=0.3$, $r_h =$1 pc, and an isothermal sphere with \(\rho
\propto r^{-2}\).

In Fig.~\ref{default} we overplot the model predictions using the
above default parameters to our data for Hydra and Centaurus. The only
parameter that is undefined in this case is the pre-enrichment level
$Z_{pre}$. We find that values of \(log_{10}(\frac{Z_{pre}}{Z_{\odot}}
)\) = $-$1.70 and $-$0.55 for Centaurus and $-$1.60 and $-$0.50 for Hydra give
the lowest deviation between the model and the data.

The default model in Fig.~\ref{default} does not agree well
with our data. The predicted (model) slope for high mass GCs is too
strong. Furthermore, the continuation of the mass-metallicity relation
towards lower masses as found by the data is not predicted by the
model. A similar conclusion was found in Mieske et al.~(\citeyear{Mieske10}) based on
the blue tilt measurement for the Fornax and Virgo clusters from HST
data. In that article we show that the tilt becomes notable one order
of magnitude in mass below the on-set mass expected by the above
default self-enrichment model. With the present (ground-based) data we
thus confirm this finding in a different environment.

In this context one should not forget that self-enrichment takes place
very early in the evolution of a star clusters. It is assumed to
happen after the explosion of the most massive stars, when their
ejecta cools down and pollutes the material of forming stars. This
happens approximately 5 Myr after the beginning of the stellar
formation process (Bailin \& Harris~\citeyear{Bailin09}). It is crucial for the model that the pollution of the ISM by SN ejecta
occurs before the bulk of (lower-mass) stars is formed.

It is thus important for the input parameters of the self-enrichment
model to represent as accurately as possible the {\it initial conditions
  of the clusters}. Turning the argument around, comparing predictions
of self-enrichment models to actual data allows us to constrain the
initial star formation conditions in globular clusters. In particular,
mass loss and the mass-radius relation of primordial GCs are very
important in shaping the present-day observed mass-metallicity
relation. We will discuss this in detail in the next two subsections.

\subsection{Inclusion of GC mass and radius evolution into the model}
\label{inclusion}
\subsubsection{Mass Loss} 
\label{sect:massloss}

           During its life time, a globular cluster will undergo mass
           loss due to stellar and dynamical evolution (e.g. Lamers et
           al.~\citeyear{Lamers05} \& \citeyear{Lamers10},
           Kruijssen~\citeyear{Kruijs09}). This accumulated mass loss
           has two effects.  

A) Based on the present-day mass as known
           from observations, the accumulated mass loss determines the
           primordial stellar cluster mass after star formation had finished. 
             This primordial mass is the value to be adopted as
             globular cluster mass $M_{GC}$ in equation \ref{f_z}. We
             reiterate that the mass of the cluster's progenitor gas
             cloud is still higher than this by a factor of
             $1/f_\star$, with $f_\star$ being the star formation
             efficiency at the time of cluster formation (equation
             \ref{f_z}).

B) Preferential loss of late dwarf stars --
           i.e. after dynamical relaxation timescales - changes the
           integrated colour of a cluster.\vspace{0.1cm}

ad A.1) Regarding stellar evolution (winds, ejecta), it is well
established that between 40\% and 60\% of the mass of the cluster is
lost after 12 Gyr (e.g. Kruijssen et al. ~\citeyear{Kruijs08}, Sippel
et al.~\citeyear{Sippel12}) because of this. We adopt a representative
mass loss of 50\% in the following.\vspace{0.1cm}

ad A.2) Mass loss due to dynamical evolution occurs when stars located in
the outskirsts of a cluster become unbound, because of their proper motion
and tidal influence of their host galaxy (e.g. Baumgardt \&
Makino~\citeyear{Baumga03}, Sippel et
al.~\citeyear{Sippel12}). Observational studies have shown that the
total dynamical mass loss experienced by a cluster can approximately
be considered as mass-independent (Jordan et al.~\citeyear{Jordan07};
see also Kruijssen~\citeyear{Kruijs09}). Theoretical arguments also
predict only a weak dependence of mass loss {\it rate} on the total
cluster mass:

\begin{equation}
\frac{dM}{dt} = \frac{M(t)^{(1-\gamma)}}{t_0}
\label{masslossrate}
\end{equation}

with $\gamma \sim 0.8$ (Lamers et al.~\citeyear{Lamers10}). Across a
factor of ten in present-day mass, the theoretically expected
accumulated mass loss would thus vary by only $\pm$30\% around the
mean. The typical accumulated mass loss for globular clusters over a
Hubble time ranges from 3 to 5 $\times (10^{5}M_\odot)$ (Jordan et
al.~\citeyear{Jordan07}). We thus adopt an accumulated dynamical mass
loss of $\Delta = 4 \times 10^{5}M_\odot$ in the following,
independent of GC present-day mass.

Combining the effects of dynamical and stellar evolution mass loss, we
thus obtain the following relation between present-day mass $M_t$ and
mass at self-enrichment $M_0$.

\begin{equation}
M_0= \frac{M_t+4\times10^5}{0.5} [\msun]
\label{eq:massloss}
\end{equation}

We note that the timescales of both contributors are quite different:
stellar evolution mass loss occurs very quickly in the early stages of
a cluster's life, while dynamical mass loss happens on very long timescales comparable to a Hubble time (e.g. Sippel et al.~\citeyear{Sippel12}).\vspace{0.1cm}

           ad B) Recent studies investigated the effect of this dynamical
           dissolussion of star clusters on their integrated colours
           (Kruijssen \& Lamers~\citeyear{Kruijs08}, Anders et
           al. ~\citeyear{Anders09}). These studies predict star
           cluster colours at a determined age and metallicity as a
           function of the ratio \(t_{dyn}\) between the current age
           and the dissolution time. It was shown that colour changes
           start to set in after $\approx$ 0.4 dissolution
           time. Qualitatively, this happens because after sufficient
           dynamical evolution, prefered stellar types are evaporated,
           namely late-type dwarfs. This in turn changes the colour of
           the stellar population bluewards (see also Mieske et
           al.~\citeyear{Mieske10}).

           We express \(t_{dyn}\) as (Jordan et
           al.~\citeyear{Jordan07}, Mieske et
           al.~\citeyear{Mieske10}):

\begin{equation}
t_{dyn} = 1 - \frac{M_{GC}}{M_{GC}+\Delta}
\end{equation}  

This allows us to calculate the colour change at each luminosity bin,
adopting the predictions by Kruijseen \& Lamers (\citeyear{Kruijs08})
in the V,I bands as a function of \(t_{dyn}\). For the present study
we then convert this to (U-I) colours with $\frac{\mathrm d
  (V-I)}{\mathrm d (U-I)} = 3.5$. This slope between both colours is
derived from the subset of our data which has U,V,I coverage.

\vspace{0.05cm}

           \subsubsection{Mass-Radius Relation}

\begin{figure}[h!]
\begin{center}
\includegraphics[width=8.6cm] {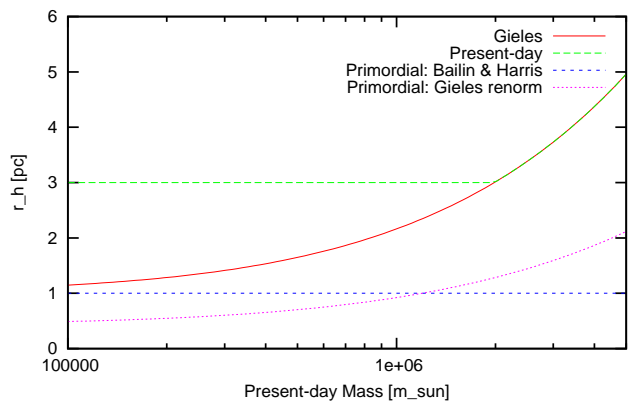}
\caption{\label{mass_radius} {\bf Red}: Star cluster mass - $r_h$
  relation from Gieles et al. (2010), normalised to have $r_h$=3pc at
  a mass of 2$\times 10^6 \msun$. {\bf Green}: present-day mass-radius
  relation (e.g. Mieske et al.~\citeyear{Mieske06}, Gieles et
  al.~\citeyear{Gieles10}): star clusters have a mass-independent
  half-light radius for masses below 2$\times 10^6 \msun$, and a
  mass-radius relation above that mass. {\bf Blue}: Primordial
  mass-radius relation (constant radius) adopted by Bailin \& Harris
  (\citeyear{Bailin09}) for star clusters at time of
  self-enrichment. {\bf Magenta}: Primordial mass-radius relation
  adopted by renormalising the red (Gieles) curve to a mean of $r_h=1
  pc$ in the observed GC mass range [$ 5\times 10^5 : 4 \times 10^6$]
  $\msun$. We note that the x-axis is the present-day mass, which is
  converted to original mass according to equation~\ref{eq:massloss}.}
\end{center}
\end{figure}

 Bailin \& Harris (\citeyear{Bailin09}) adopted for their default
 model a mass independent half-light radius \(r_h\) of 1 parsec,
 smaller than present-day average radii of 3 pc due to the early
 expansion after gas removal (e.g. Baumgardt \& Kroupa
 \citeyear{Baumga07}). The assumption of a mass-independent primordial
 GC radius may, however, be too simple, as suggested by Gieles et
 al. (\citeyear{Gieles10}). These authors argue that a primordial
 mass-radius relation down to low GC masses is dynamically more
 plausible. We will thus investigate how the predictions for the
 self-enrichment model change between a fixed radius and a mass radius
 relation for the GCs at time of self-enrichment. For the latter we
 adopt from Gieles et al. (\citeyear{Gieles10}):

\begin{equation}
log(\frac{r_{h}}{1 pc}) = -3.63 + 0.615*log(\frac{M_{0}}{M_\odot})
\label{mass_radius_form}
\end{equation}

where $r_{h}$ and $M_{0}$ are the half-mass radius and mass at time
of self-enrichment. This relation is displayed in
Figure~\ref{mass_radius}. With the given offset -3.63, one obtains a
radius of 3 pc for a $2*10^6 M_\odot$ mass cluster (Bailin \&
Harris\citeyear{Bailin09}). In Fig.~\ref{mass_radius} we also indicate
in green an approximation for the present-day mass radius relation of
GCs (e.g. Gieles et al.~\citeyear{Gieles10}, Mieske et
al.~\citeyear{Mieske06}, Murray~\citeyear{Murray09}): a piecewise
function with a mass-independent radius of 3 pc for $M<2*10^6
M_\odot$, and a mass-radius relation according to
equation~\ref{mass_radius_form} for $M>2*10^6 M_\odot$. The blue line
in Fig.~\ref{mass_radius} indicates the fixed $r_h = 1$ pc adopted by
Bailin \& Harris (2009).

      \subsection{Best fitting models}
\subsubsection{Constant vs. mass-dependent radius}

      Equations~\ref{f_z} and ~\ref{fracZ} show that the determining
      parameters of the self-enrichement efficiency of a cluster are:
      the star formation efficiency $f_\star$, the mass-radius
      relation, the density profile {\bf$\rho \propto r^{-\beta}$} and the pre-enrichment
      level. Here we aim at constraining these parameters with our data. 

\begin{figure*}
\begin{center}
\includegraphics[width = 8.6cm] {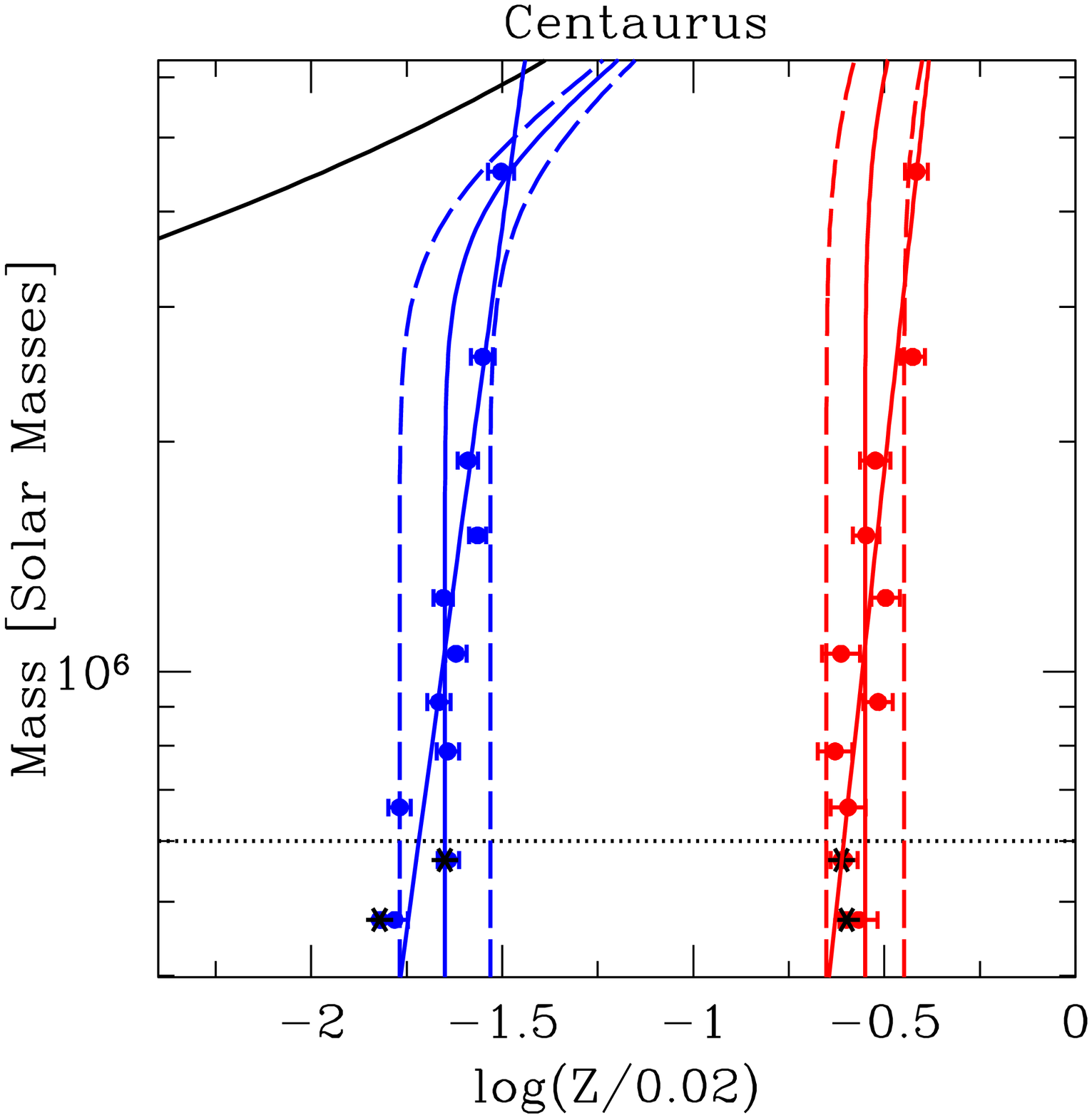}
\includegraphics[width = 8.6cm] {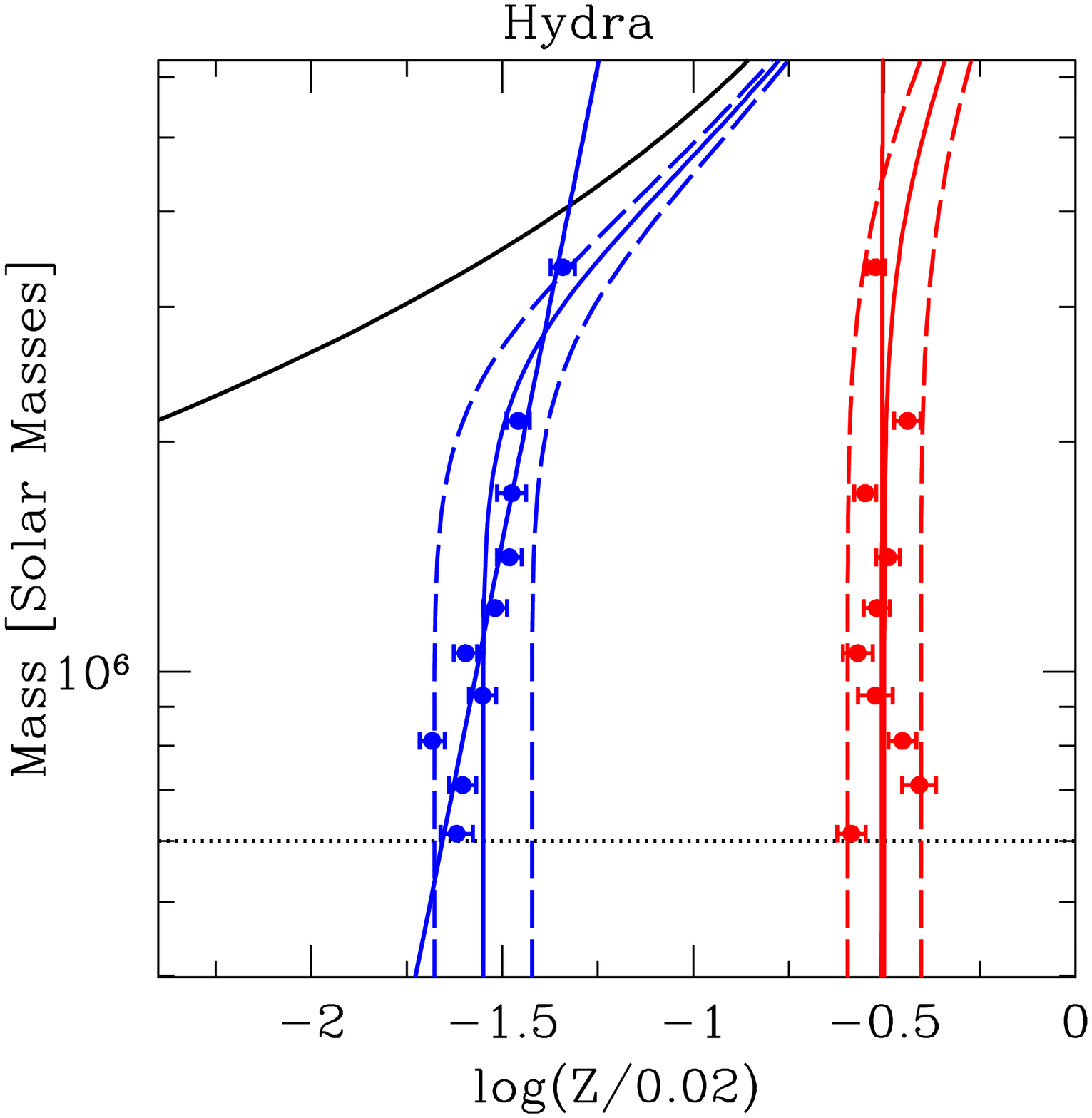}\vspace{0.3cm}\\
\includegraphics[width = 8.6cm] {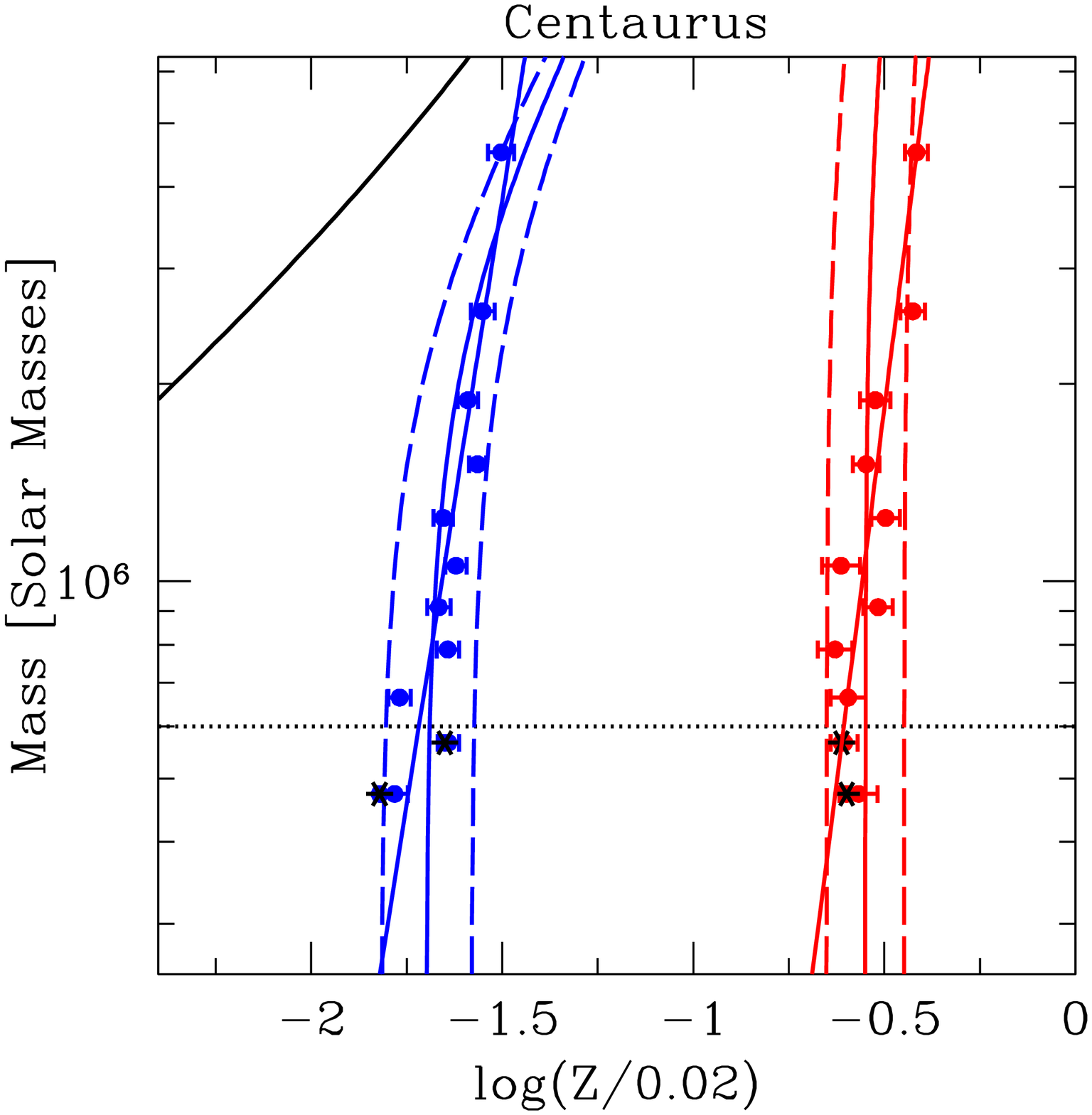}
\includegraphics[width = 8.6cm] {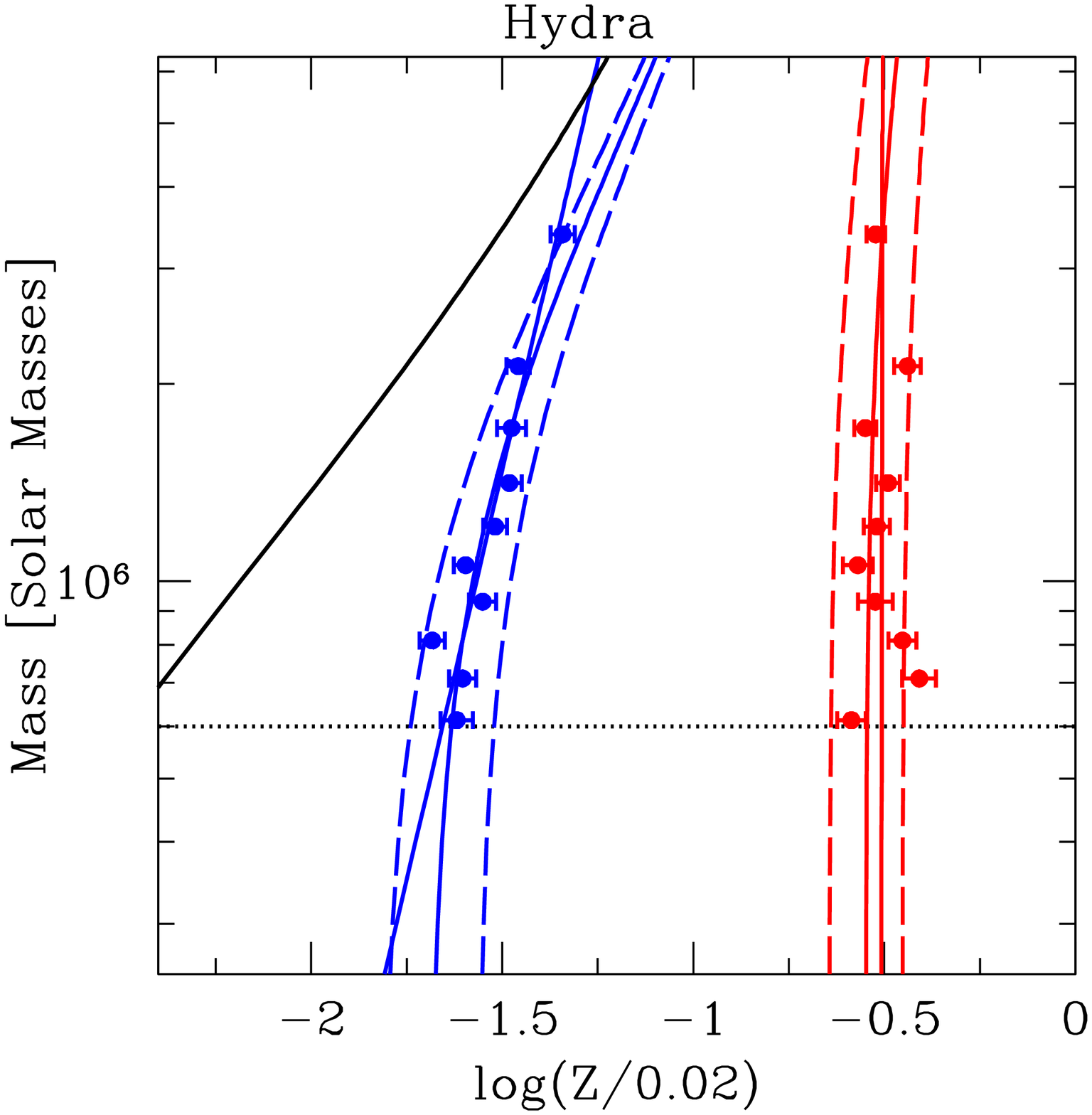}
\end{center}
\caption{\small{Model fits to our mass-metallicity data. {\bf Left:}
    Centaurus; {\bf Right:} Hydra. Compared to the model in
    Fig.~\ref{default}, we include stellar evolution mass loss and
    dynamical mass loss according to equation~\ref{eq:massloss}, and
    then determine the best-fitting star formation efficiency. The
    dotted horizontal line represents the critical mass below which
    the clusters will undergo colour changes due to dynamical mass
    evolution. Below this line, asterisks indicate the data points
    without colour changes. {\bf Top panels:} Primordial half-light
    radius $r_h$ is fixed at 1 pc. We find best-fit star formation
    efficiencies of 0.65 for Centaurus and 0.50 for Hydra and
    pre-enrichment values of (-1.65,-0.55) for Centaurus and
    (-1.55,-0.50) for Hydra. {\bf
      Bottom panels:} The mass-radius relation as in
    equation~\ref{mass_radius_form}, normalised to a mean $r_h$=1 pc
    across the GC present-day mass range [$5 \times 10^5 : 4 \times
      10^6$] $\msun$. We find best-fitting star formation efficiencies
    of 0.42 for Centaurus and 0.36 for Hydra, and pre-enrichment
    values of (-1.70,-0.55) for both Hydra and Centaurus.}}
\label{compare}
\end{figure*}

In Fig.~\ref{compare} we fit the modified model discussed in the
previous subsection to the data. Compared to the default model shown
in Fig.~\ref{default}, we keep the surface density distribution of an
isothermal sphere ($\beta=2$), and adopt the following modifications
for the other parameters:

\begin{itemize}
\item Top panels (left Centaurus, right Hydra): dynamical and stellar
  evolution mass loss is included according to equation
  ~\ref{eq:massloss}. As in the default model, a mass-independent
  primordial half-light radius of $r_h=1 pc$ is assumed. Then we vary
  the {\it star formation efficiency} and {\it pre-enrichment levels}
  to obtain the best-fit of the model to the data, using least-squares
  minimisation. We find best-fit star formation
    efficiencies of 0.65 for Centaurus and 0.50 for Hydra and
    pre-enrichment values of (-1.65,-0.55) for Centaurus and
    (-1.55,-0.50) for Hydra. The rms difference between model and data
    [Fe/H] is 0.031 dex for Centaurus and 0.028 dex for Hydra. 
\item Bottom panels (left Centaurus, right Hydra): as in the top
  panels, dynamical and stellar evolution mass loss is included
  according to equation ~\ref{eq:massloss}. In contrast to the top
  panel, we now adopt a mass-radius relation as in
  equation~\ref{mass_radius_form}, normalised to a mean $r_h$=1 pc
  across the considered GC mass range [$5 \times 10^5 : 4 \times
    10^6$] $\msun$. We find best-fitting star formation efficiencies
    of 0.42 for Centaurus and 0.36 for Hydra, and pre-enrichment
    values of (-1.70,-0.55) for both Centaurus and Hydra. The rms difference between model and data [Fe/H] is 0.020
    dex for Centaurus and 0.10 dex for Hydra.
\end{itemize}
\noindent From Fig.~\ref{compare} one notes that the introduction of a
(physically well motivated) mass-radius relation leads to a better fit
to the data, and also to more realistic star formation
efficiencies. The rms between model and data is reduced by about a
factor of two when including the mass-radius relation (bottom panel of
Fig.~\ref{compare}). With a mass-radius relation at an average $r_h$
of 1 pc, self-enrichment sets in at lower masses but increases less
rapidly towards larger masses, when compared to a mass-independent
$r_h=1$ pc.

\vspace{0.2cm}

\begin{figure*}
\begin{center}
\includegraphics[width = 8.8cm] {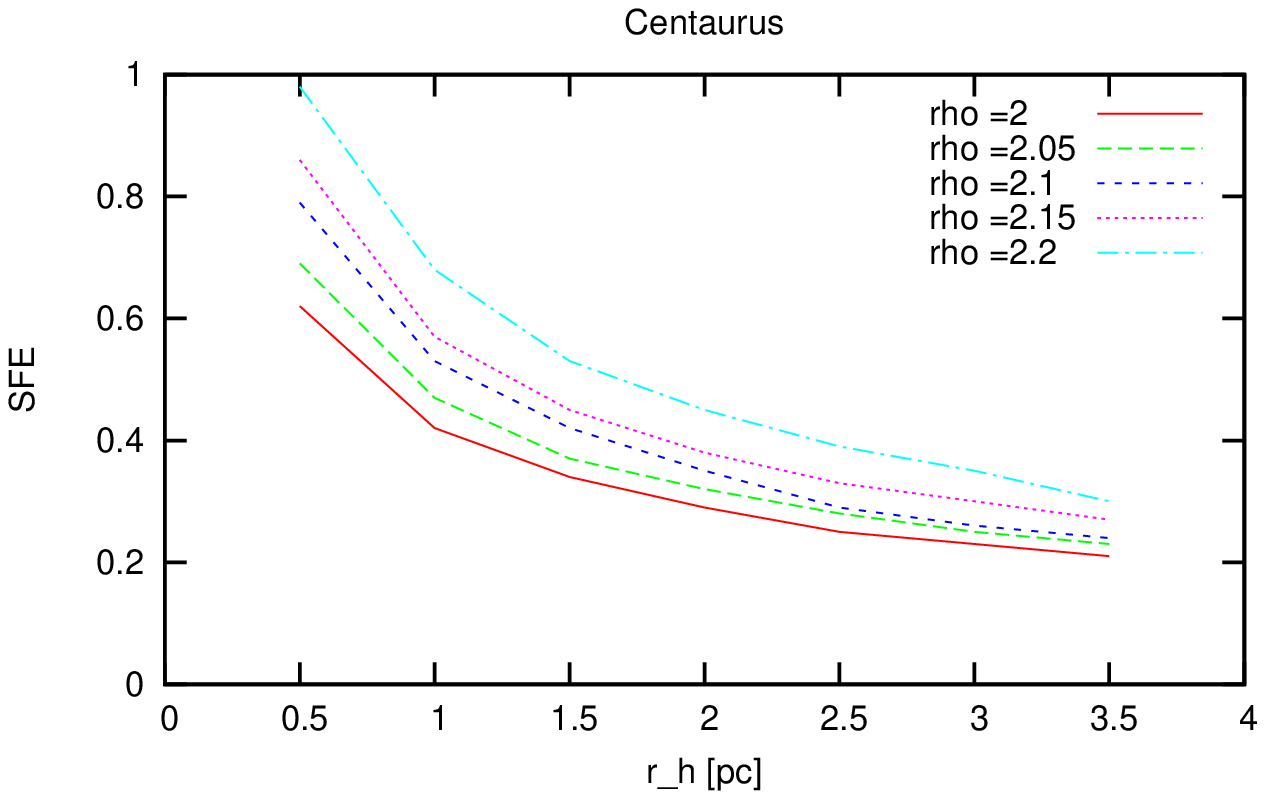}
\includegraphics[width = 8.8cm] {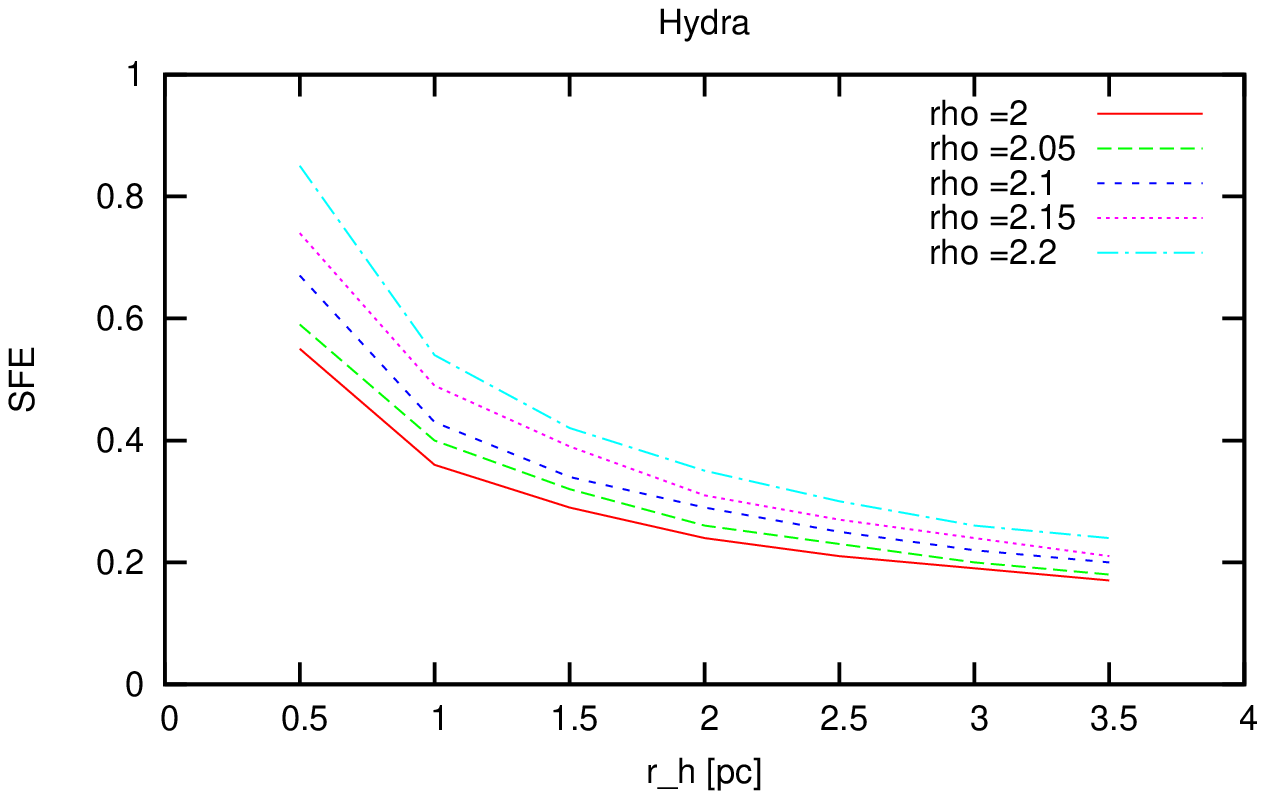}\\
\includegraphics[width = 8.8cm] {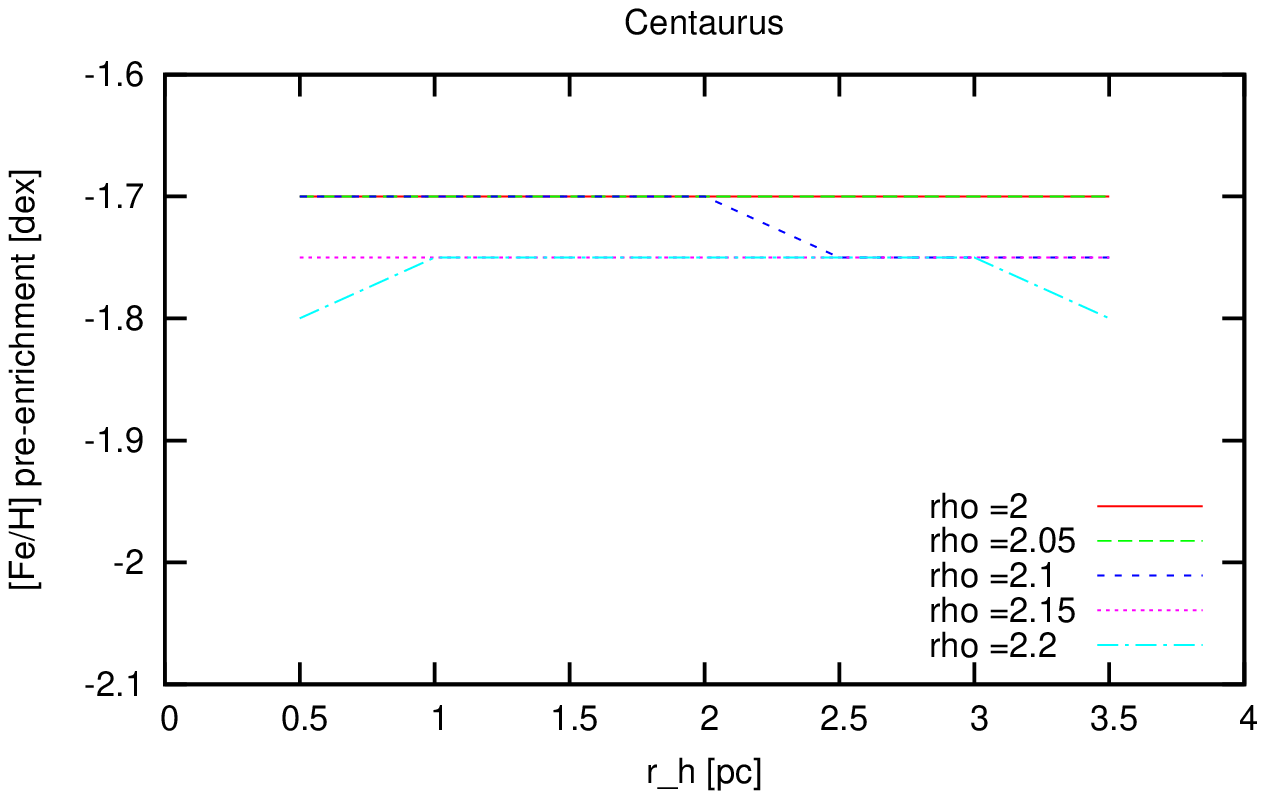}
\includegraphics[width = 8.8cm] {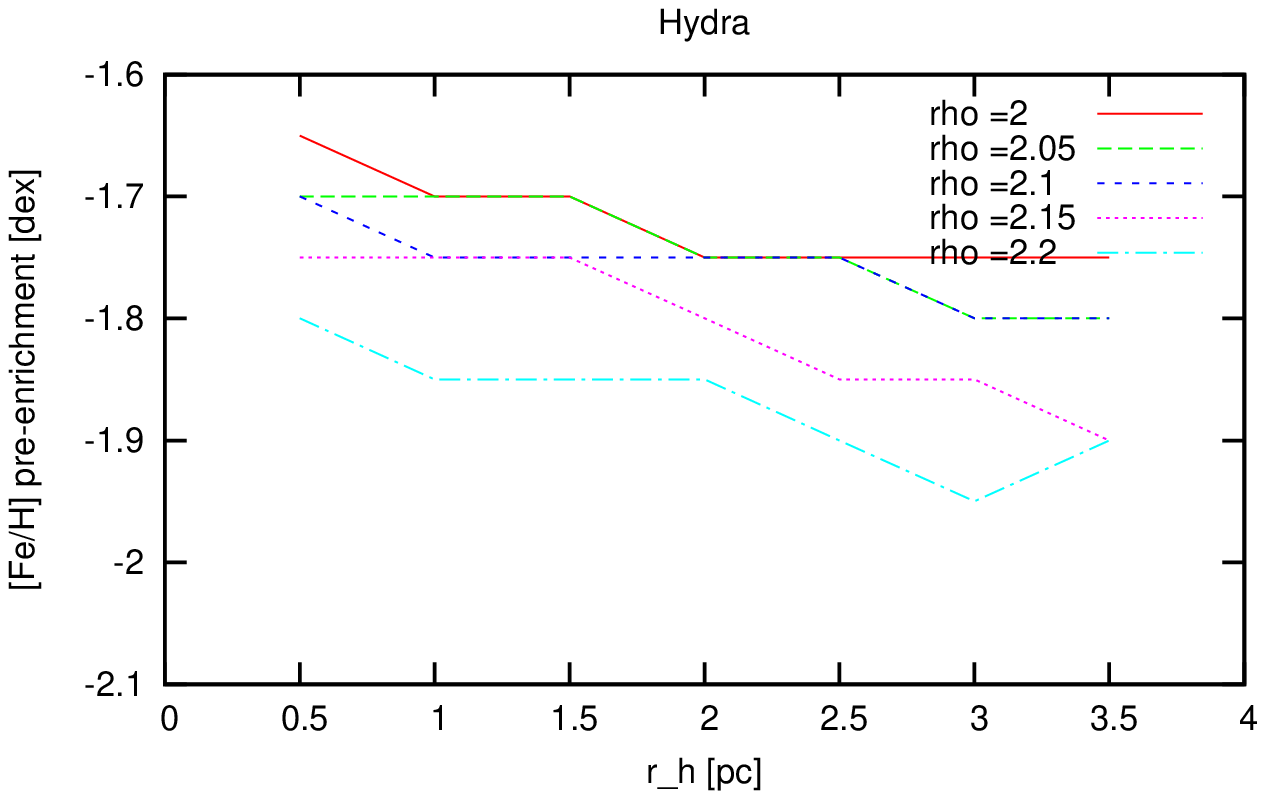}
\end{center}
\caption{\small{{\bf Top:} Best fitting star formation efficiency
    $f_\star$ for the given mean radius $r_h$ and surface density
    profile slopes $\beta$. Primordial boundary conditions as discussed
    in the text would imply $f_\star \lesssim 0.4$ and $r_h \lesssim
    2$ pc. {\bf Bottom:} Best fitting pre-enrichment level for the
    same range in mean radius $r_h$ and surface density profile slopes
    $\beta$ as above. A physically motivated lower limit for the
    pre-enrichment level is about $\sim -1.8$ dex, see text for
    details.}}
\label{compare_all}
\end{figure*}

\subsubsection{Degeneracy between radius, star-formation efficiency, surface density profile, pre-enrichment level}

\noindent In Fig.~\ref{compare_all} we provide a more general overview
of the degeneracy between self-enrichment model parameters average
radius, star-formation efficiency, surface density profile, and
pre-enrichment level. All figures assume the mass-radius relation in
equation~\ref{mass_radius_form} normalised to a mean radius within the
range [$5 \times 10^5 : 4 \times 10^6$] $\msun$ as given on the
x-axis. We first discuss the face-value constraints obtained by the
model fits, and then add physically motivated constraints to several
model parameters.

The top panels plot the best fitting star formation efficiency as a
function of average $r_h$ and for a range of surface density profile
parameter $\beta=[2.0;2.2]$. The star formation efficiency required to
match the data decreases with increasing radius, and increases with
increasing $\beta$. These trends become clear in the context of the
self-enrichment scenario: a lower star formation efficiency results in
a larger primordial cluster mass at fixed present-day mass, and thus a
deeper potential well. Self-enrichment was thus more efficient at a
given present-day mass for lower primordial star formation
efficiency. At the same time, a larger primordial radius gives a
weaker primordial self-enrichment at a fixed present-day mass given
the resulting shallower potential well.  Finally, a steeper surface
density profile (larger $\beta$) gives a deeper potential well and thus
a stronger self-enrichment.

The bottom panels show the best fitting pre-enrichment level as a
function of average $r_h$ and for a range of surface density profile
parameter $\beta=[2.0;2.2]$.

In both the upper and lower panels, the rms of the data around the
models vary by $\lesssim \pm 20\%$ among the various model flavours,
significantly less variation than the factor of two difference in rms
to the model flavour with a fixed primordial radius. That is, changing
from a fixed radius to a mass-radius relation is the most important
model update. 

For Hydra all plotted models span a total range of [0.007;0.009] dex
in rms between data and models. This very small range of $\pm$10\%
shows that no combination of the three model parameters is strongly
preferred. For Centaurus the variation in rms between model
flavours is somewhat larger with $\pm 20\%$ and a range of
[0.014;0.022]. Best-fitting models are those at large $\beta$,
i.e. very steep initial surface density profile. At a fixed $r_h=1$
pc, the rms is 0.020 for $\beta=2.0$ and 0.015 for $\beta=2.2$. At a
given $\beta$, models with smallest $r_h=0.5$ pc have about 20\% higher
rms than models with the largest $r_h=3.5$ pc.

Given the quite moderate distinction in rms between the various
best-fitting model flavours in Fig.~\ref{compare_all}, it is
appropriate to consider physically motivated boundary conditions
for the model parameters.

{\bf 1. Pre-enrichment level}: The somewhat low values of formally
possible pre-enrichment levels down to [Fe/H]$\sim -1.9$ for Hydra) in
Fig.~\ref{compare_all} are found because our data are restricted to
GCs brighter than the turn-over magnitude. The data allow reasonable
model fits for the case that self-enrichment sets in at masses lower
than our faint luminosity limit. The model curves would in this case
not straighten out at the cutoff mass of our investigation, but
continue to extend towards lower metallicities for lower masses. To
avoid the fitting degeneracy due to this extrapolation, we consider
the well-known metallicity distribution of GCs in the Milky Way as an
anchor point for the pre-enrichment level. According to the catalogue of
Harris et al. (\citeyear{Harris96}), the median metallicity of the
metal-poor GC population in the Milky Way is about [Fe/H]=-1.6
dex. Allowing for a U-I colour scale uncertainty of 0.10 mag in our
data (see Sect.~\ref{Uband}) and taking into account the slope of
$\sim$2 between [Fe/H] and U-I (equation ~\ref{colmetblue}), we thus
adopt a pre-enrichment lower limit of [Fe/H]=-1.8 dex. This excludes
for the Hydra cluster models that have average $r_h >$ 2 pc and
surface density profiles steeper than $\beta=2.1$. For the Centaurus
cluster, this restriction of pre-enrichment level to $\leq -1.8$ dex
does not constrain the range of fitting models.

{\bf 2. Star formation efficiency}: It is widely assumed that star
formation efficiency even in dense cores of molecular clouds is
$\lesssim 0.3$ (e.g. Lada \& Lada~\citeyear{Lada03}). We thus adopt a
conservative upper limit of f$_\star \leq 0.4$.

{\bf 3. Primordial half-light radius}: the typical present-day
half-light radii of most globular clusters are $\sim$3-3.5 pc (e.g. Jordan
et al.~\citeyear{Jordan05}). At the same time, primordial cluster expansion due to
gas expulsion increases the half-light radius by at least a factor of
2 (Baumgardt \& Kroupa~\citeyear{Baumga07}, Gieles et
al.~\citeyear{Gieles10} Fig. 2). We thus adopt a (still conservative)
upper limit of $r_h \lesssim 2$pc.

For the Hydra cluster, the above physical constraints restrict the
models to an isothermal sphere -- or only slightly steeper profile --
and a primordial radius $r_h$ in the range [1:2] pc, corresponding to
a range in f$_\star$ of [0.36:0.24]. The upper limit in radius
translates to the lower limit of $\sim$0.24 in f$_\star$, and
analogously the lower radius limit comes from the imposed upper limit
of $f_\star$.

For the Centaurus cluster, the star formation effiency and the $r_h<2$
limit constrain the models (not so the limit on pre-enrichment). This
restricts the models to a range of [1.5:2] pc and an isothermal
sphere, with a corresponding $f_\star$ range [0.32:0.27].

From Fig.~\ref{compare_all} it also becomes clear that the steeper
blue tilt for Hydra in comparison to Centaurus can be explained either
by a $\sim$30\% smaller average $r_h$ in Hydra at fixed $f_\star \sim
0.3$, or analogously by a $\sim$20\% smaller $f_\star$ at fixed $r_h
\sim 1.5$pc.

\subsection{Difference between Centaurus and Hydra}



It was shown above that the amplitude of the blue tilt in Hydra is
slightly stronger than in Centaurus, see Table~\ref{table_slope} and
Fig.~\ref{mmr}. The slope $\gamma=d(U-I)/dI$ is $-0.081 \pm$ 0.012 for
Hydra and $\gamma=-0.057 \pm 0.011$ for Centaurus, which formally is a
1.5$\sigma$ level difference.

\begin{figure}[]
\begin{center}
\includegraphics[width = 8.6cm] {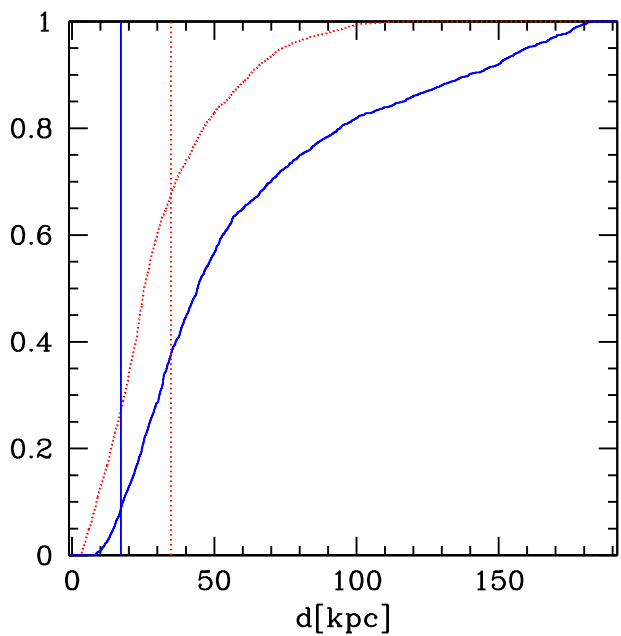}
\end{center}
\caption{\small{}Cumulative radial distribution of GC candidates in
  Hydra (dotted red line) and Centaurus (solid blue line) around the
  central galaxies NGC 3311 and NGC 4696, respectively. We note the more
  central concentration of the Hydra GCs which is imposed by the
  observational setup, see also Fig.~\ref{map}. We furthermore
  indicate as vertical lines the effective radii of the respective
  central galaxies.}
\label{cumrad}
\end{figure}

To investigate this in more detail, we show in Fig.~\ref{cumrad} the
cumulative radial distribution of GC candidates in both Hydra and
Centaurus, centered on NGC 3311 and NGC 4696 respectively. It is clear
from this figure that the GC distribution is more centrally
concentrated for Hydra than for Centaurus. We note that this is
imposed mainly by the observational setup (Fig.~\ref{map}),
which was focused on NGC 3311 in Hydra, whereas the Centaurus
observations focused on the area between the central galaxy NGC 4696
and NGC 4709. The median projected distance to NGC 3311 for the Hydra
GCs is about 25 kpc, while the median distance to NGC 4696 for the
Centaurus {\bf GCs} is about 42 kpc. The difference becomes even more
pronounced when comparing this to the respective effective radii of
both central galaxies determined from the V-band FORS data (Misgeld et
al.~\citeyear{Misgel08} \& \citeyear{Misgel09}). The effective radius
of the Centaurus cluster central galaxy NGC 4696 is 82$"$ or 17 kpc,
while for the Hydra central galaxy NGC 3311 it is 150$"$ and thus 35
kpc. Thus the galaxy half-light radius in Hydra comprises 2/3 of all
GCs, but only 10\% of the Centaurus GCs in our data set. 

A uniform area coverage in both clusters would be needed to tell
whether the GC density profiles are intrinsically different, or,
whether this higher concentration in Hydra is entirely due to the
different spatial coverage of the existing data. Analogously,
  such a uniform area coverage would help to corroborate the formally
  significant red tilt observed in Centaurus, i.e. the non-zero slope
  in the red sequence. A significant red tilt is generally not seen in
  other environments (e.g. Table~\ref{table_slope} and Harris et
  al.~\citeyear{Harris09a}). Given our lacking coverage in the central
  Centaurus field, the red sequence is only sparsely populated and
  thus the red tilt for Centaurus would need to be confirmed with more
  photometric coverage in the cluster center.




In Mieske et al. (\citeyear{Mieske10}) it was shown that within the
joint Fornax and Virgo cluster GC system, the blue tilt is more
pronounced for GCs at smaller projected radii from their host
galaxies. One may thus ask the question whether the weaker tilt in
Centaurus is related to its less centrally concentrated GC sample
compared to Hydra. As mentioned above for the Centaurus sample the
median projected distance to the central galaxy NGC 4696 is about 42
kpc. Adopting such a limit of 42 kpc between inner and outer sample,
the inner slope is $\gamma=-0.061 \pm 0.016$, and the outer slope is
$\gamma=-0.045 \pm 0.018$, marginally shallower than the inner
sample. Similarly for Hydra, with a limit of 25 kpc between inner and
outer sample, the inner slope is $\gamma=-0.061 \pm 0.016$, and the
outer slope is $\gamma=-0.045 \pm 0.018$. This qualitatively confirms
the finding in Mieske et al. (\citeyear{Mieske10}) that the blue tilt
is stronger at smaller cluster centric distance. The difference
between the inner Centaurus slope and overall Hydra slope is
insignificant, and the formally lowest difference is found between the
inner Centaurus and outer Hydra sample (-0.061 vs. -0.068).  As noted
in the previous subsection, a spatial variation in self-enrichment
efficiency can be explained by a corresponding variation in primordial
star formation efficiency or cluster radius.

In concluding we reiterate that in this study we assume a coeval
sample of old globular clusters with luminosity weighted ages around a
Hubble time. In future studies it would be worthwhile to address to
which extent the presence of young-to-intermediate age GCs may
influence the measured blue tilt, in particular its spatial variation.


\subsection{Comparison to the Milky Way Globular Clusters}

It is interesting to check whether the self-enrichment model proposed
to explain the blue tilt is consistent with what is observed in GCs of
our own Milky Way.

As noted earlier on, the cluster-to-cluster variation in [Fe/H] is
always superposed on the effect of [Fe/H] increase due to
self-enrichment within individual clusters. The statistical
significance of the trend is between 7-10$\sigma$ in our large samples
of about 2500 GCs in each environment. In the Milky Way there are only
about 50 GCs in the same absolute magnitude range (Harris et
al.~\citeyear{Harris96}). It is thus clear that no significant blue
tilt is expected to be notable in the Milky Way (see also Strader et
al. 2009), since the GC-to-GC scatter in mean [Fe/H] supersedes the
smooth and slow effect of self-enrichment.

However, the {\it width of the stellar [Fe/H] distribution} within
individual globular clusters is a much more direct tracer of possible
self-enrichment. The typical width $\sigma$ of the [Fe/H] distribution
in lower-mass GCs of the Milky Way is about 0.03 dex, see Fig.~1 of
Willman \& Strader (\citeyear{Willma12}). Then, for the most massive
GCs like Omega Centauri and M54, this [Fe/H] width increases notably
to about 0.2 dex. Such broad [Fe/H] distributions, which are also seen
in dwarf spheroidals, have often been interpreted as results of
self-enrichment. An interesting question in this context is whether
the broadened metallicity distribution for GCs like Omega Centauri
(Johnson \& Pilachowski~\citeyear{Johnso10}) is naturally explained
within a self-enrichment scenario of star clusters. Or, whether such
strong self-enrichment as observed in Omega Centauri implies that it
was once embedded in a much more massive halo (e.g. Hilker \&
Richtler~\citeyear{Hilker00}, Gnedin et al.~\citeyear{Gnedin02}). See
for this also the discussion in Willman \& Strader
(\citeyear{Willma12}).

For comparing our best-fit self-enrichment models to the GCs in the
Milky Way, we thus convert the model predictions of mean [Fe/H] to a
predicted width of the stellar [Fe/H] distribution. Such a conversion
can be very complex since individual enrichment histories are variable
-- in the Bailin \& Harris model, the metallicity spread is
  generated by the range of formation times of the low-mass stars,
  which are a few Myrs. Here we adopt a simple approach of defining
the expected width $\sigma_{\rm [Fe/H]}$ of the stellar [Fe/H]
distribution as being equal to the difference between the model
predicted [Fe/H]$_{enrich}$ and the pre-enrichment level
[Fe/H]$_{pre-enrich}$, including a minimum floor of 0.03 dex.

\begin{equation}
\sigma_{\rm [Fe/H]} = \sqrt{([Fe/H]_{enrich} - [Fe/H]_{pre-enrich})^2 + 0.03^2}
\end{equation}

\begin{figure}[]
\begin{center}
\includegraphics[width = 8.6cm] {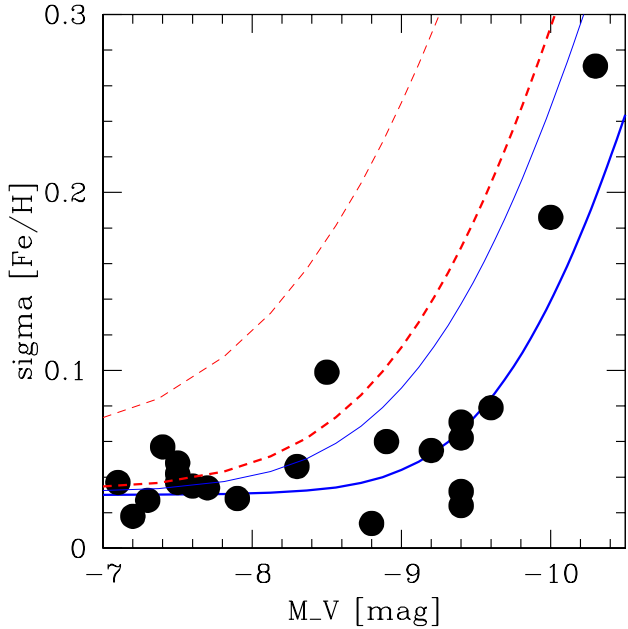}
\end{center}
\caption{\small{}Filled circles indicate the width $\sigma_{\rm
    [Fe/H]}$ of the stellar metallicity distribution in Galactic GCs
  plotted vs. their absolute visual magnitude $M_V$, compiled from
  Willman \& Strader (\citeyear{Willma12}) \& da Costa et
  al. (\citeyear{daCost14}). The blue curve indicates the
  width $\sigma_{\rm [Fe/H]}$ predicted from the best-fitting self
  enrichment model of the Centaurus cluster GCs, the red (dashed)
  curve represents the predictions from the best-fit model of the
  Hydra cluster GCs. The bold curves correspond to average primordial
  $r_h=1$ pc, the finer ones to $r_h=1.5$ pc.}
\label{gcscatter}
\end{figure}

It is important to keep in mind that the model predicted
[Fe/H]$_{enrich}$ is very close to the [Fe/H] derived from the
actual data of the Centaurus/Hydra GCs. In Fig.~\ref{gcscatter} we
compare the predicted $\sigma_{\rm [Fe/H]}$ for both the Centaurus
(blue) and Hydra (red) best-fit models for a mean $r_h=1$ (fat lines)
and $r_h=1.5$ pc (thin lines) to the $\sigma_{\rm [Fe/H]}$ observed in
Milky Way GCs (Willman \& Strader~\citeyear{Willma12}). Two things can
be noted.

1. The best-fit Centaurus model fits the Milky Way distribution quite
well, though being slightly below the Milky Way values. For $r_h=1$ pc
the sharp rise of $\sigma_{\rm [Fe/H]}$ around $M_V=-9.5$ mag to
values around 0.1-0.2 dex in the luminosity range of omega Centauri
and M54 is reproduced well. The (stronger) Hydra blue tilt implies a
larger $\sigma_{\rm [Fe/H]}$ slightly above to what is seen in the
Milky Way. We note that the only difference at $r_h=1$ pc between the
Hydra and Centaurus models is the star formation efficiency: 0.42 for
Centaurus and 0.36 for Hydra (see Fig.~\ref{compare}).

2. The curves for $r_h=1$ pc imply a smaller $\sigma_{\rm [Fe/H]}$
spread than the curves at $r_h=1.5$ pc. This is because a larger
primordial radius requires a lower pre-enrichment level to fit the
data, and thus the measured metallicities at high masses are farther
away from the pre-enrichment levels. In this context we note that {\it
  at the same} pre-enrichment level, variations in $r_h$ and star
formation efficiency are degenerate as already mentioned above. For
example, the bold curve for Centaurus can be represented both by
combination of an average primordial radius $r_h=1$ pc and SFE=0.42,
{\it and} an average primordial radius $r_h=3$ pc and SFE=0.23. The
latter example may fit better the case of omega Centauri with its
present day $r_h=8$pc.

The above example shows that broadenend metallicity distributions like
found in some massive MW globular clusters may be natural consequences
of cluster self-enrichment processes, without the need to invoke an
additional embeddening in a more massive halo (e.g. a dwarf galaxy) at
time of formation (e.g. Dinescu et al.~\citeyear{Dinesc99}, Majewski
et al.~\citeyear{Majews00}, Gnedin et al.~\citeyear{Gnedin02}, Bekki
et al.~\citeyear{Bekki06}).

One caveat is that the self-enrichment model applied here (Bailin \&
Harris~\citeyear{Bailin09}) assumes quasi-instantaneous reprocessing
of SNII ejecta, i.e. on timescales that are short compared to dynamical timescales. This is a simplifying assumption that would need to be
elaborated in further detail, given that an age spread of several
billion years has been reported for the stellar populations of omega
Centauri (e.g. Hilker et al.~\citeyear{Hilker04}).

\section{Summary and Conclusions}
\label{summary}
We analyse the colour-magnitude relation of globular clusters (the
'blue tilt') in the central 100 kpc of the Hydra and Centaurus galaxy
clusters. This analysis is based on deep FORS1 photometry in the U and
I bands, which provides a very broad wavelength baseline and thus high
metallicity sensitivity. The final sample comprises about 2500 GC
candidates in each cluster down to $M_I \sim -9$ mag, half a magnitude
brighter than the GC luminosity function turnover. \\

We obtain the following results

\begin{itemize}

\item In both clusters we measure a signficiant 'blue tilt', i.e. a
  colour-magnitude relation for the blue globular cluster
  subpopulation. We find $\gamma = d(U-I)/d(I) = $-0.057 $\pm$ 0.011
  for Centaurus and $\gamma=-0.081 \pm 0.012$ for Hydra. We confirm
  previous findings that the blue tilt already sets in at present-day
  masses well below $10^6 \msun$.

\item We convert colours and luminosities to mass and metallicity
  using TERAMO models (Raimondo et al.~\citeyear{Raimon05}), assuming
  old GC ages of 11-13 Gyrs. We thus obtain a mass-metallicity scaling
  of $Z \propto M^{0.27 \pm 0.05}$ for Centaurus and $Z \propto
  M^{0.40 \pm 0.06}$ for Hydra. This is the same range than found in
  previous literature measurements of the GC mass-metallicity scaling,
  investigating different environments and mostly using HST data.

\item We compare the measured mass-metallicity dependence of GCs on star cluster self-enrichment model predictions using the model of
  Bailin \& Harris (\citeyear{Bailin09}). To use a realistic
  primordial star cluster mass at the time of self-enrichment, we
  include both stellar evolution and dynamical mass loss effects. We
  find that the model fits are significantly improved (reducing rms by
  a factor of two with respect to the data) when assuming a primordial
  mass-radius relation of star clusters according to Gieles et
  al. (\citeyear{Gieles10}), instead of a fixed, mass independent,
  cluster radius.

\item We investigate and illustrate the degeneracy between model input
  parameters average radius $r_h$, star formation efficiency
  $f_\star$, pre-enrichment level, and surface density profile
  steepness. The best fit within physically motivated boundary
  conditions is obtained for $r_h \sim 1-1.5$ pc, $f_\star \sim 0.3 -
  0.4$, a pre-enrichment level of $[Fe/H] ~ -1.7$ and a surface
  density profile of an isothermal sphere. The slightly steeper blue
  tilt for Hydra can be explained either by a $\sim$30\% smaller
  average $r_h$ at fixed $f_\star \sim 0.3$, or analogously by a
  $\sim$20\% smaller $f_\star$ at fixed $r_h \sim 1.5$ pc.

\item We note that the GC sample in the Centaurus cluster is less
  centrally concentrated than the Hydra sample, given the
  observational setup of the observed fields. We show that both in
  Hydra and Centaurus the blue tilt is stronger for GCs with smaller
  projected distances to their host galaxies. We thus argue that the
  slightly stronger tilt in Hydra could be due to its more centrally
  concentrated GC sample.

\item We find that the $M_V$ vs. $\sigma_{\rm [Fe/H]}$ distribution of
  Galactic Globular Clusters is consistent with the GC self-enrichment
  trend observed for the Centaurus cluster.

\end{itemize}

We conclude that 

\begin{enumerate}
\item U-I photometry allows us to accurately measure colour-magnitude
relation of GCs

\item GC self-enrichment sets in at present-day masses well below
  $10^6 \msun$

\item The adoption of a primordial mass-radius relation in the
  self-enrichment model of Bailin \& Harris (\citeyear{Bailin09})
  allows us to fit data very well without requiring non-canonical star
  formation efficiency, surface density profile, or average radii.

\item The comparison between the blue tilt strength of Hydra and
  Centaurus supports previous findings of variations of
  self-enrichment efficiency with cluster-centric distance. Spatial
  variations of primordial cluster radius or star formation
  efficiencies can explain such a behaviour.

\item Broadened metallicity distribution like found in some MW
  globular clusters can be natural consequences of cluster
  self-enrichment, without the need to invoke an additional
  embeddening in a more massive halo (e.g. a dwarf galaxy) at time of
  formation.

\end{enumerate}

\begin{acknowledgements}
We thank the anonymous referee for very constructive comments which helped to improve this paper. JF acknowledges financial support from the ESO Chile Office for Science.

\end{acknowledgements}

\bibliographystyle{aa}
\bibliography{Fensch_accepted}

\end{document}